\documentclass[10pt,aps,twocolumn,showpacs,amsmath,amssymb,superscriptaddress,longbibliography,prb]{revtex4-2}
\usepackage{physics}
\usepackage{graphicx}
\usepackage{epsf}
\usepackage{dcolumn}
\usepackage{amsmath}
\usepackage{cases}
\usepackage[dvipsnames]{xcolor}
\usepackage{soul}

\makeatletter
\let\old@makecaption=\@makecaption
\usepackage{subcaption}
\let\@makecaption=\old@makecaption
\makeatother
\usepackage{blindtext}
\usepackage[normalem]{ulem}
\usepackage{booktabs}
\usepackage{tikz}
\usetikzlibrary{arrows.meta,positioning,calc}
\usepackage{multirow}
\usepackage{array}

\usepackage[]{url}
\usepackage{bbold}
\usepackage[colorlinks, linkcolor=blue, citecolor=red]{hyperref}
\usepackage[capitalise]{cleveref}

\renewcommand{\b}[1]{{\boldsymbol{#1}}}

\newcommand{\nn}{\nonumber}

\usepackage{cases}

\makeatletter
\@booleanfalse\titlepage@sw
\makeatother

\begin{document}

\title{Induction of $p$-wave and $d$-wave order parameters in $s$-wave superconductors with light pulses}
\author{Hennadii Yerzhakov}
\affiliation{Nordita, Stockholm University, and KTH Royal Institute of Technology, Hannes Alfvéns väg 12, SE-106 91 Stockholm, Sweden}
\author{Alexander Balatsky}
\affiliation{Nordita, Stockholm University, and KTH Royal Institute of Technology, Hannes Alfvéns väg 12, SE-106 91 Stockholm, Sweden}
\affiliation{Department of Physics, University of Connecticut, Storrs, Connecticut 06269, USA}

\date\today

\begin{abstract}
 We construct a generalized time-dependent Ginzburg–Landau model to demonstrate the possibility of inducing $p$- and $d$-wave components in an originally pure $s$-wave centrosymmetric superconductor via microwave radiation. In this framework, specializing to $O_h$ point-group symmetry, we introduce gradient terms that couple the $s$-wave superconducting order parameter with other symmetry-allowed components. The singlet-to-singlet gradient terms are quadratic in spatial derivatives, while, in the presence of spin–orbit coupling, linear-in-derivatives terms coupling singlet and triplet order parameters are also permitted. Through the minimal substitution procedure, these terms enable coupling between different superconducting order parameters via the vector potential, thereby leading to the generation of $p$-wave, $d$-wave, and other symmetry-allowed components. Such a manipulation of the superconducting state locally via a microwave beam could be considered as one more facet of the concept of quantum printing.
\end{abstract}

\maketitle

\section{Introduction}

The study of the interaction of superconducting (SC) condensate with electromagnetic (EM) radiation has a long history. 
Recent theoretical efforts have been directed towards exploration of the possibility of inducing superconducting correlations of lower symmetries (including triplet correlations) in originally $s$-wave superconductors (SCs) by microwave or sub-THz radiations~\cite{gassner2024,yokoyama2025,fu2025}.

In Ref.~\cite{gassner2024}, based on a mix of microscopic and Ginzburg-Landau approaches, a theoretical proposal of switching the singlet $s$-wave superconducting state to a metastable triplet state via exposure to EM radiation is presented. The key ingredients in that paper are spin-orbit coupling and irradiation by an electromagnetic wave that breaks the inversion symmetry dynamically. In this work, we generalize the symmetry-based Ginzburg-Landau theory formulation and do not require dynamical inversion symmetry breaking as a necessary mechanism for inducing triplet correlations via EM radiation. In Ref.~\cite{yokoyama2025}, an induction of the triplet correlation in an $s$-wave SC with SOC with EM waves is studied within the Floquet-Magnus expansion. However, the models used in that work are for a non-centrosymmetric SC with Rashba-type spin-orbit coupling, which generically allows singlet-triplet admixture. 

In this work, we construct a time-dependent GL theory concentrating on systems with $O_h$ point group symmetry, in which the $s$-wave order parameter (OP) is coupled to other symmetry-allowed singlet and triplet OPs via gradient terms. The key point of this work is the formulation of the Lifshitz-type invariant~\cite{landau1969statistical} (first order in spatial derivatives), which is allowed in the presence of SOC, coupling $s$- and $p$-wave SC OPs. Under EM irradiation, this coupling induces a dynamical $p$-wave component, whereas other gradient couplings generate additional singlet components. 

Such dynamically induced singlet-triplet states might potentially be a platform for a Floquet-engineered topological superconductivity, induction of a transient topological superconducting state, and ultimately can lead to an alternative route to topological SC using light/microwave pulses to produce a $p$-wave SC. We can see that this route can have advantages as it does not require heterostructures with attendant disorder and imperfection effects. The proposed manipulation of the superconducting state with light represents another facet of quantum printing~\cite{yeh2025a,yeh2025b,aeppli2025quantumprinting}, in which the spatiotemporal structure of the light gauge potential modifies the superconducting state.

Although the present work focuses on the Lifshitz-type invariant in the presence of a time-dependent vector potential, a possible implication in the static vector potential case is the emergence of triplet correlations in the vortex state of an originally $s$-wave superconductor at zero applied magnetic field, possibly allowing for vortex fractionalization. This coupling may also be relevant for the explanation of the magnetic memory effect in 4Hb-TaS$_2$~\cite{Persky2022,mandal2025}. 

The paper is structured as follows: In~\cref{sec:transformation}, we recapitulate how the superconducting gap transforms under symmetry operations. In~\cref{sec:model}, we construct a generalized time-dependent Ginzburg-Landau model for the $O_h$-symmetric crystal introducing gradient terms that couple the $s$-wave OP with the triplet $p$-wave and other singlet OPs. In~\cref{sec:tdglequations}, we derive equations of motion for the formulated model. In~\cref{sec:results}, we present some analytical and numerical results on the induced superconducting components. Finally, we conclude in~\cref{sec:conclusions}.

\section{Transformation properties of the superconducting gap}
\label{sec:transformation}

The SC gap function for a single-band SC can generically be written as~\cite{SigristUeda1991}
\begin{align}
\label{eq: gap function}
    \hat{\Delta}(\b{k}) = i (\sigma_0 \psi(\b{k}) + \b{d}(\b{k})\cdot\b{\sigma}) \sigma_y,
\end{align}
where $\sigma_0$ is the $2\times2$ identity matrix and $\b{\sigma} = (\sigma_x, \sigma_y, \sigma_z)$ is the vector of Pauli matrices. The terms with $\psi(\b{k})$ and $\b{d}(\b{k})$ correspond to the singlet and triplet pairings, respectively. The group of symmetries is $\mathcal{G} = G \times SU(2) \times \mathcal{T} \times U(1)$ in the absence of SOC. This is reduced to $\mathcal{G} = G^D \times \mathcal{T} \times U(1)$ in the presence of SOC. $G$ is the space group of the lattice (we will concentrate on the corresponding point group symmetry only), $G^D$ is its double group and $\mathcal{T}$ is the time-reversal symmetry. At the highest transition temperature in a centrosymmetric superconductor, either the singlet or the triplet component is nonzero. However, at lower temperatures, the ground state may exhibit singlet--triplet mixing~\cite{Annett1990,SigristUeda1991,Sigrist2005,Normand1994,ODonovan1995,Sorensen1991,HutchinsonMarsiglio2020,SenarathPramodh2025}.

In the absence of SOC, under the symmetry operations, the creation operators(in momentum space) transform as
\begin{align}
    g_O c_{\b{k},s}^\dagger &= c_{D^{-}(g_O) \b{k},s},\ g_S c_{\b{k}}^\dagger = D_{1/2}(g_S) c_{{\b{k}}},\\ \nn  
    \mathcal{T} c_{\b{k}} &= -i \sigma_y c_{-\b{k}},\ \Phi c_{\b{k},s} = e^{i \phi} c_{\b{k},s},
\end{align}
where $g_O,g_S$, and $\Phi$ are members of the symmetry groups $G, SU(2),$ and $U(1)$, respectively. $D^{-}(g)$ denote matrices in the three-dimensional odd representation of the group $G$. $D_{1/2}$ is the two-dimensional spin-$1/2$ representation of $SU(2)$. Therefore, their actions on the gap function are

\begin{align}
    g_O \hat{\Delta}(\b{k}) &= \hat{\Delta}(D^-(g_O)\b{k}),\ \\ \nn
    g_S \hat{\Delta}(\b{k}) &= D_{1/2}(g_S)^T \hat{\Delta}(\b{k}) D_{1/2}(g_S),\\ \nn
    \mathcal{T} \hat{\Delta}(\b{k}) &= \sigma_y \hat{\Delta}(-\b{k})^* \sigma_y,\ \Phi \hat{\Delta}(\b{k}) = e^{2i\phi} \hat{\Delta}(\b{k}).
\end{align}

These operations can be rewritten for $\psi(\b{k})$ and $\b{d}(\b{k})$ in~\cref{eq: gap function}:
\begin{align}
    g_O \psi(\b{k}) &= \psi(D^-(g_O)\b{k}),\ g_O \b{d}(\b{k}) = \b{d}(D^-(g_O) \b{k}),\\ \nn
    g_S \psi(\b{k}) &= \psi(\b{k}),\ g_S \b{d}(\b{k}) = D^+(g_S) \b{d}(\b{k}),\\ \nn
    \mathcal{T} \psi(\b{k}) &= \psi(-\b{k})^*,\ \mathcal{T} \b{d}(\b{k}) = -\b{d}(-\b{k})^*,\\ \nn
    \Phi \psi(\b{k}) &= e^{2 i\phi}\psi(\b{k}),\ \Phi \b{d}(\b{k}) = e^{2 i\phi} \b{d}(\b{k}),
\end{align}
where $D^{+}(g)$ are matrices in the even three-dimensional representation of $G$.

The presence of SOC ``locks" rotations in spin and orbital spaces. Consequently, in this case 
\begin{align}
    g_{OS} \hat{\Delta}(\b{k}) = D_{1/2}^G(g_{OS})^T \hat{\Delta}(D^-(g_{OS})\b{k}) D_{1/2}^G(g_{OS}),
\end{align}
where $g_{OS} \in G^D$, $D_{1/2}^G$ is a $2\times2$ spinorial irrep of the double group $G^D$, and we instead have
\begin{align}
    g_{OS} \psi(\b{k}) &= \psi(D^-(g_{OS}) \b{k}),\\ \nn
    \ g_{OS} \b{d}(\b{k}) & = D^+(g_{OS}) \b{d}(D^-(g_{OS})\b{k}).
\end{align}

We want to investigate what couplings through the electromagnetic vector potential between OPs of different symmetries are possible, which are the allowed gradient terms in the GL theory. 

\section{Model}
\label{sec:model}

\subsection{SO(3) symmetry, no SOC}
For continuous rotational symmetry, and even $l$, the gap function can be expanded in the basis functions of the $\mathcal{D}_l$ representation as
\begin{align}
    \hat{\Delta}(\b{k}) = \sum_{m=-l}^l c_{m} Y_{lm}(\b{k}) i \sigma_y = \sum_{m=-l}^l c_{m} \hat{\Delta}_m,
\end{align}
and for odd $l$
\begin{align}
    \hat{\Delta} = \sum_{m=-l}^l \sum_{\hat{n}=\hat{x},\hat{y},\hat{z}} c_{mn} Y_{lm}(\b{k}) i (\b{\sigma}\cdot \hat{n}) \sigma_y,
\end{align}

Thus, the $s$-wave is characterized by one coefficient $\hat{\Delta}_s(\b{k}) = s\hat{\Delta}_s(\b{k})$, and the $p$-wave by nine coefficients
\begin{align}
    \hat{\Delta}_p = \sum_{i=1}^3 \sum_{\alpha=\hat{x},\hat{y},\hat{z}} P_{i\alpha} \hat{\Delta}_{i\alpha}
\end{align}

Transferring the transformation properties to the coefficients, we want to form an invariant that couples $s$ and $P_{i\alpha}$ and contains derivatives that transform according to the $\mathcal{D}_1$ representation, i.e., their transformations are performed by applying regular SO(3) rotation matrices. We schematically (without specifying indices and numerical coefficients) write such a putative invariant and how it transforms as
\begin{align}
    I_{\grad} =s\cdot\b{\grad}^n P^m \xrightarrow[g_O,g_S]{}s\cdot {D^-}^n(g_O)\b{\grad}^n {D^-}^m(g_O) \cdot \\ \nn{D^+}^m(g_S)  P^m.
\end{align}
Since $D_\pm$ are simply rotation matrices for proper rotations, any polynomial constructed from them can be reduced to a constant by appropriately pairing indices and inserting Levi--Civita tensors $\epsilon_{ijk}$, using $R^{T}R=\mathbb{1}$ and $\det R=1$, where $R$ is a rotation matrix.
Thus, to form an invariant, we have to have an even number of $D^-(g_O)$ matrices to satisfy invariance under inversion and $m\geq 2$ to satisfy invariance under spin rotations. Within the TDGL model, such invariants provide terms of at least linear order in $P$ to the TDGL equations, which will be zero as long as the $p$-wave components are zero. Therefore, such terms will be unable to induce the triplet component via microwave irradiation or via non-uniformity of the $s$-wave OP in the vortex state. 

Note that couplings between singlet OPs of different symmetries, e.g., between $s$- and $d$-wave components, are allowed in second order in spatial derivatives and in first order in the OPs even without SOC.

\subsection{SO(3) symmetry, SOC is present}
The situation is different in the presence of SOC. After SOC is turned on, the full group $SO(3)\times SU(2)$ reduces to $SU(2)$ (acting in combined spin and orbital space). Correspondingly, the irreducible representation, according to which the SC gap transformed in the case of no SOC, becomes reducible representations of $SU(2)$. 
Generically, under the SOC, the physical spin seizes to be a good quantum number for labeling energy-degenerate single-particle states, from which Cooper pairs are formed. Therefore, we switch to the appropriate pseudo-spin basis, in which the single-particle Hamiltonian is diagonalized. In this basis, the singlet and triplet components of the superconducting gap retain their symmetry-defined meaning under inversion, whereas in the physical spin basis the singlet and triplet components generally mix. 
The eigenbasis of the SC gap for even $l$ now transforms according to
\begin{align}
    \mathcal{D}_l \otimes \mathcal{D}_0 = \mathcal{D}_l,
\end{align}
and for odd $l$ as
\begin{align}
    \mathcal{D}_l \otimes \mathcal{D}_1 = \mathcal{D}_{l-1} \oplus \mathcal{D}_l \oplus \mathcal{D}_{l+1}.
\end{align}
Thus, for the $p$-wave component with $l=1$, it is possible to form an invariant of the first order in both the spatial derivatives and the $p$-wave components $P_{i\alpha}$.

Specifically, such an invariant transforms under a $g_{OS}$ operation as 
\begin{align}
    &I_{\grad} =a_{ij\alpha}s^*\partial_i P_{j\alpha} +c.c. \xrightarrow[g_{OS}]{} \\ \nn
    &a_{ij\alpha} s\cdot {D_{ii'}^-}(g_{OS})\partial_{i'} {D_{jj'}^-}^m(g_{OS}) \cdot {D_{\alpha \alpha'}^+}^m(g_{OS})  P_{j' \alpha'} + c.c.,
\end{align}
where $a_{ij\alpha}$ are the coefficients to be determined.
Since $D^- = D^+ \cdot i$, where $i=\pm1$ depending on whether the operation is proper or improper rotation, and $\det D^+ =1$ can be rewritten in the form 
\begin{align}
    \epsilon_{ijk} D_{ii'}^+ D_{jj'}^+ D_{kk'}^+ = \epsilon_{i'j'k'},
\end{align}
where $\epsilon_{ijk}$ is the totally antisymmetric tensor, it is easy to see that $a_{ij\alpha} \sim \epsilon_{ij\alpha}$. Given that $I_{\grad}$ is integrated over the volume in the Lagrangian, using integration by parts, it can be rewritten in an antisymmetric form
\begin{align}
    I_{\grad} \sim \epsilon_{ij\alpha}\left( s^*\partial_i P_{j\alpha} - (\partial_i s^*) P_{j\alpha} +c.c.\right).
\end{align}
An important feature of this Lifshitz-like term is that it introduces terms of zeroth order in $P_{i\alpha}$ to the equation of motions, which leads to the generation of the triplet component via application of periodic (electromagnetic wave) or static (constant magnetic field) vector potentials. The strength of coefficients $a_{ij\alpha}$ is expected to be proportional to the strength of the SOC in the first approximation. Linear in gradient terms of this type have been discussed in the context of non-centrosymmetric superconductors~\cite{mineev1994helical,samokhin2004,mineev2008,samokhin2013gradient,kapustin2022} and multiband superconductors~\cite{kamatani2022,nagashima2024,dunbrack2025}.
For single-band centrosymmetric SCs, it was originally introduced in Ref.~\cite{mineev1994helical} within the GL framework, where the focus of study was on helical SC states, and to our best knowledge, not discussed in the literature after that. In addition, while assumed, the importance of SOC was not discussed there. It recently reappeared in Ref.~\cite{gassner2024}, where it was microscopically derived for a specific model on a honeycomb lattice with Kane-Mele spin-orbit coupling~\cite{kanemele2005}.

\subsection{$O_h$ symmetry with SOC}
We proceed by constructing a theory for the case of cubic symmetry with inversion center, $O_h$, and non-zero SOC. $O_h$ has five even and five odd representations, with the corresponding basis functions given in~\cref{tab: Basis functions}. The gap function in the $\Gamma$ representation can be written as
\begin{align}
    \hat{\Delta} = \sum_{i=1}^{d_\Gamma} \eta_\Gamma^i \hat{\Delta}_\Gamma^i,
\end{align}
where $d_\Gamma$ is the dimension of the representation $\Gamma$.

\begin{table*}[!htbp]
\centering
\caption{Irreducible representations and their corresponding basis functions $\hat{\Delta}_\Gamma^i$~\cite{SigristUeda1991}. For even irreps, $\hat{\Delta}_\Gamma^i = i\sigma_y \psi_\Gamma(\b{k})$, and for odd irreps $\hat{\Delta}_\Gamma^i = i (\b{\sigma}\cdot\b{d}_\Gamma^i(\b{k})) \sigma_y$}
\setlength{\tabcolsep}{8pt}
\renewcommand{\arraystretch}{1.2}
\begin{tabular}{@{}l l l l@{}}
\toprule
\textbf{Irrep $\Gamma$} & \textbf{Basis functions} & \textbf{Irrep $\Gamma$} & \textbf{Basis vector-functions} \\
\midrule

$A_{1g}$ & $\psi_{A_{1g}}(\mathbf{k})=1,\;k_x^2+k_y^2+k_z^2$ &
$A_{1u}$ & $d_{A_{1u}}(\mathbf{k})=\hat{\mathbf{x}}k_x+\hat{\mathbf{y}}k_y+\hat{\mathbf{z}}k_z$ \\

$A_{2g}$ & $\psi_{A_{2g}}(\mathbf{k})=(k_z^2-k_y^2)(k_y^2-k_x^2)(k_z^2-k_x^2)$ &
$A_{2u}$ & $\begin{aligned}[t]
d_{A_{2u}}(\mathbf{k})={}&\hat{\mathbf{x}}k_x(k_z^2-k_y^2)+\hat{\mathbf{y}}k_y(k_x^2-k_z^2)\\
&+\hat{\mathbf{z}}k_z(k_y^2-k_x^2)
\end{aligned}$ \\

$E_{g}$ & $\begin{aligned}[t]
\psi_{E_{g}}^1(\mathbf{k}) &= 2k_z^2-k_x^2-k_y^2,\\
\psi_{E_{g}}^2(\mathbf{k}) &= \sqrt{3}(k_x^2-k_y^2)
\end{aligned}$ &
$E_{u}$ & $\begin{aligned}[t]
d_{E_{u},
}^1(\mathbf{k}) &= 2\hat{\mathbf{z}}k_z-\hat{\mathbf{x}}k_x-\hat{\mathbf{y}}k_y,\\
d_{E_{u}}^2(\mathbf{k}) &= \sqrt{3}(\hat{\mathbf{x}}k_x-\hat{\mathbf{y}}k_y)
\end{aligned}$ \\

$T_{1g}$ & $\begin{aligned}[t]
\psi_{T_{1g}}^1(\mathbf{k}) &= k_y k_z(k_y^2-k_z^2),\\
\psi_{T_{1g}}^2(\mathbf{k}) &= k_z k_x(k_z^2-k_x^2),\\
\psi_{T_{1g}}^3(\mathbf{k}) &= k_x k_y(k_x^2-k_y^2)
\end{aligned}$ &
$T_{1u}$ & $\begin{aligned}[t]
d_{T_{1u}}^1(\mathbf{k}) &= \hat{\mathbf{y}}k_z-\hat{\mathbf{z}}k_y,\\
d_{T_{1u}}^2(\mathbf{k}) &= \hat{\mathbf{z}}k_x-\hat{\mathbf{x}}k_z,\\
d_{T_{1u}}^3(\mathbf{k}) &= \hat{\mathbf{x}}k_y-\hat{\mathbf{y}}k_x
\end{aligned}$ \\

$T_{2g}$ & $\begin{aligned}[t]
\psi_{T_{2g}}^1(\mathbf{k}) &= k_y k_z,\\
\psi_{T_{2g}}^2(\mathbf{k}) &= k_z k_x,\\
\psi_{T_{2g}}^3(\mathbf{k}) &= k_x k_y
\end{aligned}$ &
$T_{2u}$ & $\begin{aligned}[t]
d_{T_{2u}}^1(\mathbf{k}) &= \hat{\mathbf{y}}k_z+\hat{\mathbf{z}}k_y,\\
d_{T_{2u}}^2(\mathbf{k}) &= \hat{\mathbf{z}}k_x+\hat{\mathbf{x}}k_z,\\
d_{T_{2u}}^3(\mathbf{k}) &= \hat{\mathbf{x}}k_y+\hat{\mathbf{y}}k_x
\end{aligned}$ \\
\bottomrule
\end{tabular}
\label{tab: Basis functions}
\end{table*}

\subsubsection{$A_{1g}$-to-triplet.}
We start by forming invariants linear in gradient, singlet ($s$-wave) $A_{1g}$ OP, and triplet OPs. Using the direct product~\cref{tab: Product table}~\cite{atkins1970tables,katzerohcharactertable} in~\cref{app:producttable}, we find that the only such coupling is between $A_{1g}$ ($s$-wave) and $T_{1u}$ ($p$-wave) irreps. Indeed, given that the components of $\grad$ transform according to $T_{1u}$ irrep, for the invariant term of the structure $(\eta_{A_{1g}})^* (\grad \eta_{\Gamma_u})$ to exist, there should be a trivial irrep in the decomposition of $A_{1g} \otimes T_{1u} \otimes \Gamma_{u}$. Using a theorem that a direct product of two real irreps contains a trivial one if only those two irreps coincide~\cite{landau1981quantum}, we see that such an invariant can be formed only if $\Gamma_u = T_{1u}$. For an $O_h$ point group, it is given by~\cref{eq: LA1gT1u}. In~\cref{tab: Invariants}, we list Lifshitz-like invariants coupling $s$-wave OP with the symmetry allowed triplet OPs for all crystallographic point groups.

\begin{table*}[!htbp]
\centering
\caption{Lifshitz-like invariants coupling $s$-wave OP with symmetry-allowed triplet OPs in different centrosymmetric crystallographic point groups.}
    \begin{tabular}{|c|c|}
\hline
Group & Lifshitz-like invariants \\
\hline

\multirow{3}{*}{\parbox[c]{4em}{\centering $C_i$}} &
$\eta_{A_g}^* \partial_x \eta_{A_u} - (\partial_x \eta_{A_g}^*) \eta_{A_u}$ + c.c.\\
& $\eta_{A_g}^* \partial_y \eta_{A_u} - (\partial_y \eta_{A_g}^*) \eta_{A_u}$ + c.c.\\
& $\eta_{A_g}^* \partial_z \eta_{A_u} - (\partial_z \eta_{A_g}^*) \eta_{A_u}$ + c.c.\\
\hline

\multirow{3}{*}{\parbox[c]{4em}{\centering $C_{2h}$}} &
$\eta_{A_g}^* \partial_x \eta_{B_u} - (\partial_x \eta_{A_g}^*) \eta_{B_u}$ + c.c.\\
& $\eta_{A_g}^* \partial_y \eta_{B_u} - (\partial_y \eta_{A_g}^*) \eta_{B_u}$ + c.c.\\
& $\eta_{A_g}^* \partial_z \eta_{A_u} - (\partial_z \eta_{A_g}^*) \eta_{A_u}$ + c.c.\\
\hline

\multirow{3}{*}{\parbox[c]{4em}{\centering $D_{2h}$}} &
$\eta_{A_g}^* \partial_x \eta_{B_{3u}} - (\partial_x \eta_{A_g}^*) \eta_{B_{3u}}$ + c.c.\\
& $\eta_{A_g}^* \partial_y \eta_{B_{2u}} - (\partial_y \eta_{A_g}^*) \eta_{B_{2u}}$ + c.c.\\
& $\eta_{A_g}^* \partial_z \eta_{B_{1u}} - (\partial_z \eta_{A_g}^*) \eta_{B_{1u}}$ + c.c.\\
\hline

\multirow{3}{*}{\parbox[c]{4em}{\centering $C_{4h}$\\ $S_6$\\ (or $C_{3i}$)}} &
$\eta_{A_g}^* (\partial_x \eta_{E_u}^1 + \partial_y \eta_{E_u}^2 ) -
\left( (\partial_x \eta_{A_g}^*) \eta_{E_u}^1 + (\partial_y \eta_{A_g}^*) \eta_{E_u}^2 \right)$ + c.c.\\
&
$\eta_{A_g}^* (\partial_x \eta_{E_u}^2 - \partial_y \eta_{E_u}^1 ) -
\left( (\partial_x \eta_{A_g}^*) \eta_{E_u}^2 - (\partial_y \eta_{A_g}^*) \eta_{E_u}^1 \right)$ + c.c.\\
& $\eta_{A_g}^* \partial_z \eta_{A_u} - (\partial_z \eta_{A_g}^*) \eta_{A_u}$ + c.c.\\
\hline

\multirow{2}{*}{\parbox[c]{4em}{\centering $D_{4h}$\\ $D_{3d}$}} &
$\eta_{A_{1g}}^* (\partial_x \eta_{E_u}^1 + \partial_y \eta_{E_u}^2 ) -
\left( (\partial_x \eta_{A_{1g}}^*) \eta_{E_u}^1 + (\partial_y \eta_{A_{1g}}^*) \eta_{E_u}^2 \right)$ + c.c.\\
& $\eta_{A_{1g}}^* \partial_z \eta_{A_{2u}} - (\partial_z \eta_{A_{1g}}^*) \eta_{A_{2u}}$ + c.c.\\
\hline

\multirow{3}{*}{\parbox[c]{4em}{\centering $C_{6h}$}} &
$\eta_{A_g}^* (\partial_x \eta_{E_{1u}}^1 + \partial_y \eta_{E_{1u}}^2 ) -
\left( (\partial_x \eta_{A_g}^*) \eta_{E_{1u}}^1 + (\partial_y \eta_{A_g}^*) \eta_{E_{1u}}^2 \right)$ + c.c.\\
&
$\eta_{A_g}^* (\partial_x \eta_{E_{1u}}^2 - \partial_y \eta_{E_{1u}}^1 ) -
\left( (\partial_x \eta_{A_g}^*) \eta_{E_{1u}}^2 - (\partial_y \eta_{A_g}^*) \eta_{E_{1u}}^1 \right)$ + c.c.\\
& $\eta_{A_g}^* \partial_z \eta_{A_u} - (\partial_z \eta_{A_g}^*) \eta_{A_u}$ + c.c.\\
\hline

\multirow{2}{*}{\parbox[c]{4em}{\centering $D_{6h}$}} &
$\eta_{A_{1g}}^* (\partial_x \eta_{E_{1u}}^1 + \partial_y \eta_{E_{1u}}^2 ) -
\left( (\partial_x \eta_{A_{1g}}^*) \eta_{E_{1u}}^1 + (\partial_y \eta_{A_{1g}}^*) \eta_{E_{1u}}^2 \right)$ + c.c.\\
& $\eta_{A_{1g}}^* \partial_z \eta_{A_{2u}} - (\partial_z \eta_{A_{1g}}^*) \eta_{A_{2u}}$ + c.c.\\
\hline

\parbox[c]{4em}{\centering $T_h$} &
$\eta_{A_{g}}^* (\partial_x \eta_{T_u}^1 + \partial_y \eta_{T_u}^2 + \partial_z \eta_{T_u}^3 ) -
\left( (\partial_x \eta_{A_{g}}^*) \eta_{T_u}^1 + (\partial_y \eta_{A_{g}}^*) \eta_{T_u}^2 + (\partial_z \eta_{A_{g}}^*) \eta_{T_u}^3 \right)$ + c.c.\\
\hline

\parbox[c]{4em}{\centering $O_h$} &
$\eta_{A_{1g}}^* (\partial_x \eta_{T_{1u}}^1 + \partial_y \eta_{T_{1u}}^2 + \partial_z \eta_{T_{1u}}^3 ) -
\left( (\partial_x \eta_{A_{1g}}^*) \eta_{T_{1u}}^1 + (\partial_y \eta_{A_{1g}}^*) \eta_{T_{1u}}^2 + (\partial_z \eta_{A_{1g}}^*) \eta_{T_{1u}}^3 \right)$ + c.c.\\
\hline
\end{tabular}
\label{tab: Invariants}
\end{table*}

\subsubsection{$A_{1g}$-to-singlet}
The couplings between $s$-wave and other singlet components need at least two gradients to form an invariant. Decomposing direct products (showing only the $A_{1g}$ irrep and using ellipses for all other terms) $A_{1g}\otimes T_{1u} \otimes T_{1u} \otimes \Gamma_{g}$, using~\cref{tab: Product table}, 
\begin{align}
    A_{1g}\otimes T_{1u} \otimes T_{1u} \otimes \begin{pmatrix} A_{1g} \\ A_{2g} \\ E_g \\ T_{1g} \\ T_{2g} \end{pmatrix} = \begin{pmatrix} A_{1g} + ... \\ ... \\ A_{1g} + ... \\ A_{1g} + ... \\ A_{1g} + ... \end{pmatrix},
\end{align}
we find that couplings to all other singlets, except for $A_{2g}$, are possible. Again, because such terms are linear in $\eta_{\Gamma_{g}}$, such invariants will yield terms independent of $\eta_{\Gamma_{g}}$ in the TDGL eqs., and thus capable of inducing those components via application of periodic or static vector potentials. In the case of the latter one, in Ref.~\cite{talkachov2025} it has been shown that subdominant $d$-wave OP might nucleate in the vicinity of the $s$-wave vortices and lead to type-$1.5$ superconductivity.

\begin{figure}[h!]
\centering
\begin{tikzpicture}[scale=2,>=latex]
  \tikzset{
    baseedge/.style={very thick,red,-},
    apexedge/.style={ultra thick,red,-},   
    centeredge/.style={very thick,blue!60!black,-},
    centeredge2/.style={ultra thick,blue!60!black,-},
    lab/.style={font=\Large},
    vertex/.style={circle,fill=black,inner sep=3pt}, 
  }

  \coordinate (Apex) at (1.7,1.35);      
  \coordinate (B1)   at (0,0);           
  \coordinate (B2)   at (3.1,-0.2);      
  \coordinate (B3)   at (1.0,-1.7);      

  \coordinate (Cent) at (1.45,-0.55);


  \draw[apexedge] (Apex) to[left=6] (B1);
  \draw[apexedge] (Apex) -- (B2);
  \draw[apexedge] (Apex) to[right=8] (B3);

  \draw[centeredge] (Cent)--(B1);
  \draw[centeredge] (Cent)--(B2);

  \draw[baseedge] (B1)--(B2);
  \draw[centeredge2] (Cent)--(Apex);
  \draw[centeredge] (Cent)--(B3);
  \draw[baseedge] (B2)--(B3);
  \draw[baseedge] (B3)--(B1);

  \node[vertex] at (Apex) {};
  \node[vertex] at (B1) {};
  \node[vertex] at (B2) {};
  \node[vertex] at (B3) {};
  \node[vertex] at (Cent) {};

  \node[lab,anchor=east]  at ($(Apex)+(+1.1,0.2)$) {$\Delta_{A_{1g}} {\scriptscriptstyle (s\mathrm{-wave})}$};
  \node[lab,anchor=east]  at ($(B1)+(0.05,-0.15)$)  {$\underset{{\scriptscriptstyle (d\mathrm{-wave})}}{\Delta_{E_g}}$};
  \node[lab,anchor=west]  at ($(B2)+(-0.0,-0.15)$)   {$\underset{{\scriptscriptstyle (g\mathrm{-wave})}}{\Delta_{T_{1g}}}$};
  \node[lab,anchor=west]  at ($(B3)+(0.12,-0.05)$) {$\Delta_{T_{2g}} {\scriptscriptstyle (d\mathrm{-wave})}$};
  \node[lab,anchor=west]  at ($(Cent)+(-0.08,-0.23)$) {$\underset{{\scriptscriptstyle (p\mathrm{-wave})}}{\Delta_{T_{1u}}}$};
\end{tikzpicture}
\caption{Schematic diagram of gradient couplings between different SC OPs. Red and blue  colors stand for the singlet-to-singlet and singlet-to-triplet couplings, respectively. Thicker lines correspond to couplings to the $s$-wave OP, which is the only non-zero OP in the absence of the vector potential.}
\label{fig: Schematic couplings}
\end{figure}
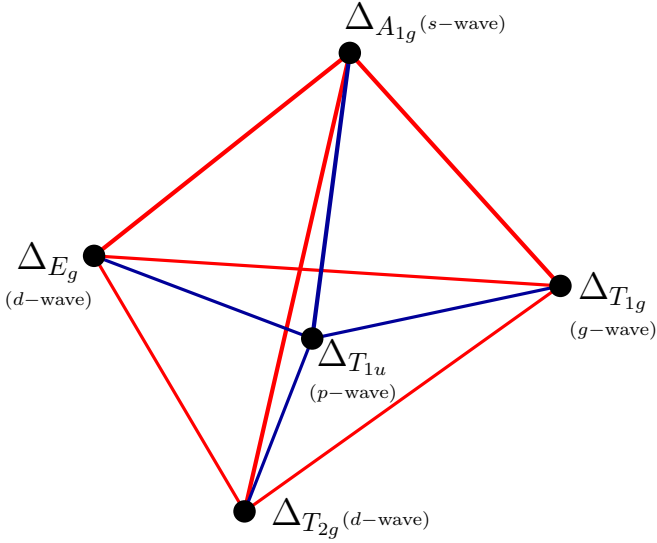

Together, the $s$-wave ($A_{1g}$) OP could be coupled via vector potential to the $T_{1u}$ ($p$-wave), $E_g$ ($d_{x^2-y^2}$ and $d_{z^2}$), $T_{1g}$ ($g$-wave), and $T_{2g}$ ($d_{xy}$, $d_{xz}$, and $d_{yz}$) OPs. We thus also investigate possible couplings between them. The schematic picture showing possible gradient couplings between different SC components in the constructed GL theory is illustrated in~\cref{fig: Schematic couplings}.

\subsection{Lagrangian}
The full static Lagrangian density is thus
\begin{align}
\label{eq: full static Lagrangian}
    \mathcal{L}_{st}= &\sum_{\Gamma \in \Gamma_r}\mathcal{L}_{\Gamma} + \sideset{}{'}\sum_{\Gamma \in \Gamma_r} \mathcal{L}_{A_{1g}\Gamma} + \mathcal{L}_{T_{1u}E_g} + \mathcal{L}_{T_{1u}T_{1g}} + \mathcal{L}_{T_{1u} T_{2g}} +\\ \nn
     &\mathcal{L}_{E_g T_{1g}} + \mathcal{L}_{E_g T_{2g}} + \mathcal{L}_{T_{1g} T_{2g}},
\end{align}
where $\Gamma_r$ is the set of $T_{1u}$ and all even irreps of the group, excluding $A_{2g}$, and the prime in the superscript of the sum means excluding $A_{1g}$ from the sum. $\mathcal{L}_\Gamma$'s are given by~\cite{SigristUeda1991}
\begin{widetext}
\begin{align}
\mathcal{L}_{A_{1g}}
=& r_{A_{1g}}\,\lvert \eta_{A_{1g}} \rvert^{2}
+ \sum_{i} a_{A_{1g}}\,\lvert D_i \eta_{A_{1g}} \rvert^{2}
+ b_{A_{1g}}\,\lvert \eta_{A_{1g}} \rvert^{4}, \\
\mathcal{L}_{T_{1u}} =&
r_{T_{1u}} \sum_{i=1}^{3}\!\left|\eta^{\,i}_{T_{1u}}\right|^{2}
+ a^{1}_{T_{1u}} \sum_{i=1}^{3}\!\left|D_i \eta^{\,i}_{T_{1u}}\right|^{2}
\\[2pt] \nn
&+ a^{2}_{T_{1u}}\Big(
|D_x \eta^{\,2}_{T_{1u}}|^{2}+|D_x \eta^{\,3}_{T_{1u}}|^{2}
+|D_y \eta^{\,1}_{T_{1u}}|^{2}+|D_y \eta^{\,3}_{T_{1u}}|^{2}
+|D_z \eta^{\,1}_{T_{1u}}|^{2}+|D_z \eta^{\,2}_{T_{1u}}|^{2}
\Big)
\\[2pt] \nn
&+ a^{3}_{T_{1u}}\Big[
(D_x\eta^{\,1}_{T_{1u}})^{*}(D_y\eta^{\,2}_{T_{1u}})
+(D_y\eta^{\,2}_{T_{1u}})^{*}(D_z\eta^{\,3}_{T_{1u}})
+(D_z\eta^{\,3}_{T_{1u}})^{*}(D_x\eta^{\,1}_{T_{1u}})
+\text{c.c.}\Big]
\\[2pt] \nn
&+ a^{4}_{T_{1u}}\Big[
(D_x\eta^{\,2}_{T_{1u}})^{*}(D_y\eta^{\,1}_{T_{1u}})
+(D_y\eta^{\,3}_{T_{1u}})^{*}(D_z\eta^{\,2}_{T_{1u}})
+(D_z\eta^{\,1}_{T_{1u}})^{*}(D_x\eta^{\,3}_{T_{1u}})
+\text{c.c.}\Big]
\\[2pt] \nn
&+ b^{1}_{T_{1u}}\!\left(\sum_{i=1}^{3}\left|\eta^{\,i}_{T_{1u}}\right|^{2}\right)^{2}
+ b^{2}_{T_{1u}}\left|\sum_{i=1}^{3} \left(\eta^{\,i}_{T_{1u}}\right)^2\right|^{2}
+ b^{3}_{T_{1u}}\Big(
|\eta^{\,1}_{T_{1u}}|^{2}|\eta^{\,2}_{T_{1u}}|^{2}
+|\eta^{\,1}_{T_{1u}}|^{2}|\eta^{\,3}_{T_{1u}}|^{2}
+|\eta^{\,2}_{T_{1u}}|^{2}|\eta^{\,3}_{T_{1u}}|^{2}
\Big)\,\\ \nn
%
\mathcal{L}_{E_g}
=& r_{E_g}\sum_{i=1}^{2}\!\left|\eta_{E_g}^{\,i}\right|^{2}
  + a^{1}_{E_g}\sum_{j=1}^{3}\sum_{i=1}^{2}\!\left|D_j\eta_{E_g}^{\,i}\right|^{2}       +b^{1}_{E_g}\!\left(\sum_{i=1}^{2}\left|\eta_{E_g}^{\,i}\right|^{2}\right)^{2}
      + b^{2}_{E_g}\!\left(\eta_{E_g}^{\,1*}\eta_{E_g}^{\,2}
-\eta_{E_g}^{\,2*}\eta_{E_g}^{\,1}\right)^{2} \\[2pt] \nn
&\quad + a^{2}_{E_g}\Big\{
  2\big(\,|D_z\eta_{E_g}^{\,2}|^{2}-|D_z\eta_{E_g}^{\,1}|^{2}\,\big)
  -\big(\,|D_x\eta_{E_g}^{\,1}|^{2}-|D_x\eta_{E_g}^{\,2}|^{2}\,\big)
  -\big(\,|D_y\eta_{E_g}^{\,1}|^{2}-|D_y\eta_{E_g}^{\,2}|^{2}\,\big) \\[-2pt] \nn
&\qquad\qquad
  + \sqrt{3}\,\big( (D_x\eta_{E_g}^{\,1})(D_x\eta_{E_g}^{\,2})^{*}
                   -(D_y\eta_{E_g}^{\,1})(D_y\eta_{E_g}^{\,2})^{*}
                   + \text{c.c.} \big)
  \Big\} \\[2pt] \nn
\mathcal{L}_{T_{1g}} = &  \mathcal{L}_{T_{1u}} \biggr\vert_{T_{1u}\rightarrow T_{1g}},\ \mathcal{L}_{T_{2g}} =  \mathcal{L}_{T_{1u}} \biggr\vert_{T_{1u}\rightarrow T_{2g}},
\end{align}
\end{widetext}
where $D_i=\partial_i - i 2e/\hbar \cdot A_i$ is the covariant derivative, and $e$ is the electron charge. Representing the terms that couple $A_{1g}$ with other irreps as the sum of the gradient and uniform parts, $\mathcal{L}_{A_{1g}\Gamma} = \mathcal{L}_{A_{1g}\Gamma}^\nabla + \mathcal{L}_{A_{1g}\Gamma}^u$, for the gradient parts, we obtain

\begin{widetext}
\begin{align}
\label{eq: LA1gT1u}
\mathcal{L}_{A_{1g},T_{1u}}^\nabla
& = a_{A_{1g}T_{1u}}
\sum_{i=1}^3\left\{
\eta^{*}_{A_{1g}}\,D_i \eta^{\,i}_{T_{1u}}
- (D_i \eta_{A_{1g}})^{*}\,\eta^{\,i}_{T_{1u}}
+ \text{c.c.}
\right\}, \\
\mathcal{L}_{A_{1g},E_g}^\nabla
& = a_{A_{1g}E_g}\!
\Big\{
\tfrac12 (D_x\eta^{\,1}_{E_g})^{*} D_x \eta_{A_{1g}}
-\tfrac12 (D_x\eta^{\,2}_{E_g})^{*} D_x \eta_{A_{1g}}
-\tfrac12 (D_y\eta^{\,1}_{E_g})^{*} D_y \eta_{A_{1g}} \\ \nn
&\quad -\tfrac12 (D_y\eta^{\,2}_{E_g})^{*} D_y \eta_{A_{1g}} + (D_z\eta^{\,2}_{E_g})^{*} D_z \eta_{A_{1g}}
+\text{c.c.}
\Big\}, \\
\mathcal{L}_{A_{1g},T_{1g}}^\nabla
& = a_{A_{1g}T_{1g}}
\Big\{
-\,D_x\eta_{A_{1g}}\,(D_y\eta^{\,1}_{T_{1g}})^{*}
+\,D_y\eta_{A_{1g}}\,(D_x\eta^{\,1}_{T_{1g}})^{*}
-\,D_x\eta_{A_{1g}}\,(D_z\eta^{\,2}_{T_{1g}})^{*} \\ \nn
&\quad +\,D_z\eta_{A_{1g}}\,(D_x\eta^{\,2}_{T_{1g}})^{*}
-\,D_y\eta_{A_{1g}}\,(D_z\eta^{\,3}_{T_{1g}})^{*}
+\,D_z\eta_{A_{1g}}\,(D_y\eta^{\,3}_{T_{1g}})^{*}
+\text{c.c.}
\Big\}, \\
\mathcal{L}_{A_{1g},T_{2g}}^\nabla
& = a_{A_{1g}T_{2g}}
\Big\{
D_x\eta_{A_{1g}}\,(D_y\eta^{\,1}_{T_{2g}})^{*}
+ D_y\eta_{A_{1g}}\,(D_x\eta^{\,1}_{T_{2g}})^{*}
+ D_x\eta_{A_{1g}}\,(D_z\eta^{\,2}_{T_{2g}})^{*} \\ \nn
&\quad + D_z\eta_{A_{1g}}\,(D_x\eta^{\,2}_{T_{2g}})^{*}
+ D_y\eta_{A_{1g}}\,(D_z\eta^{\,3}_{T_{2g}})^{*}
+ D_z\eta_{A_{1g}}\,(D_y\eta^{\,3}_{T_{2g}})^{*}
+ \text{c.c.}\Big\},
\end{align}
\end{widetext}

and for the uniform parts, $\mathcal{L}_{A_{1g}\Gamma}^u$, we use
\begin{align}
    \mathcal{L}_{A_{1g}\Gamma}^u =& c_{A_{1g}\Gamma}^1 \abs{\eta_{A_{1g}}}^2 \sum_{i=1}^{d_\Gamma} \abs{\eta_{\Gamma}^i}^2 \\ \nn
    &+ c_{A_{1g}\Gamma}^2 \left(  \eta_{A_{1g}}^2 \sum_{i=1}^{d_\Gamma} (\eta_{\Gamma}^{i*})^2 + \eta_{A_{1g}}^{*2} \sum_{i=1}^{d_\Gamma} (\eta_{\Gamma}^i)^2 \right),
\end{align}
where $d_\Gamma$ is the dimension of the $\Gamma$ irrep. When $\Gamma = T_{1u}$, there are no other invariants quartic in OPs. In particular, there are no invariants cubic in $\eta_{T_{1u}}$ and linear in $\eta_{A_{1g}}$ since the inversion symmetry requires an even number of ungerand irreps entering the invariant. When $\Gamma$ is gerund, there are invariants cubic in $\eta_{\Gamma}$ and linear in $\eta_{A_{1g}}$. However, we do not include them here, as they are of higher order in $\eta_\Gamma$. The rest of the terms entering into~\cref{eq: full static Lagrangian} are given in~\cref{app:restofthelagrangian}.

We further generalize this static Lagrangian model to a time-dependent GL theory. For $s$-wave superconductors, the simplest TDGL model is obtained by extending the static GL equations to~\cite{kopnin2001theory}
\begin{align}
\label{eq: s-wave TDGL}
    \Gamma_s D_t \eta_s = -\frac{\delta F}{\delta \eta_s^*},
\end{align}
where $F=\int d^d x \mathcal{L}_{st}$ is the free energy, $D_t = \partial_t + i 2e/\hbar \cdot \phi$ is a covariant time derivative, in which $\mu$ is an electric potential, and $\Gamma_s$ is the relaxation time. It is
microscopically justified  for gapless superconductors close to $T_c$. A generalized TDGL model was derived microscopically for dirty gapped $s$-wave superconductors near $T_c$~\cite{kramer1978theory,watts1981nonequilibrium}. ~\cref{eq: s-wave TDGL} can be obtained by introducing the Rayleigh functional~\cite{landau1982mechanics}
\begin{align}
\label{eq: s-wave TDGL Lagrangian}
    R = \int d^dr \Gamma_s \abs{D_t \eta_s}^2.
\end{align}
We do not derive the TDGL model, corresponding to the static Lagrangian in~\cref{eq: full static Lagrangian},  microscopically. Instead, we use generalized~\cref{eq: s-wave TDGL Lagrangian}:
\begin{align}
\label{eq: TDGL Lagrangian}
    R =& \int d^d r \sum_{\mu \in \Gamma_r} \sum_{\substack{i,j=1\\ i\neq j}}^{d_\mu} \Gamma_\nu^{ij} (D_t \eta_\nu^i)^* (D_t \eta_\nu^j) + \\ \nn 
    &\int d^d r \sum_{\substack{\mu, \nu \in \Gamma_r \\ \mu \neq \nu}} \sum_{i=1}^{d_\mu} \sum_{j=1}^{d_\nu} \Gamma_{\mu \nu}^{ij} (D_t \eta_\mu^i)^* (D_t \eta_\nu^j).
\end{align}
To preserve rotational invariance of the time-derivative terms, we have to set $\Gamma_{\mu \nu}^{ij} = 0$, and $\Gamma_\nu^{ij} = \Gamma_\nu \delta_{ij}$.

\section{TDGL equations}
\label{sec:tdglequations}

\subsection{Equations of motion}

The equations of motion are given by
\begin{align}
\frac{\delta \mathcal{L}}{\delta \eta_\nu^{i\,*}} - \partial_j \frac{\delta \mathcal{L}}{\delta (\partial_j \eta_\nu^{i\,*})} + \frac{\delta R}{\delta (\partial_t \eta_\nu^{i\,*})}=0.
\end{align}

We explicitly write down these equations in dimensionless units. The dimensionless coordinate, time, vector potential, and current are: 
\begin{align}
\b{r}' = \b{r}/\xi, t'=t/\tau_0, \b{A}'=\b{A}/A_0, \b{j}' = \b{j}/j_0, \phi'=\phi/\phi_0,
\end{align}
where $\xi$ is the coherence length of the pure $s$-wave uniform SC, $\tau_0 = \mu_0 \sigma \lambda^2$, where $\mu_0$ is the vacuum permeability, $\sigma$ is the normal conductivity, $\lambda$ is the magnetic penetration depth of the pure $s$-wave SC, $A_0=\xi B_{c2}$, where $B_{c2} = \Phi_0/(2 \pi \xi^2)$ is the upper critical magnetic field of the pure $s$-wave SC, $j_0 = \xi B_{c2}/(\mu_0 \lambda^2)$, and $\phi_0 = \xi j_0/\sigma$. The dimensionless GL parameters are
\begin{align}
&\Gamma_\nu' = \frac{\Gamma_\nu}{\tau_0 (-r_{A_{1g}})},\, r_{\Gamma_\nu}'=\frac{r_{\Gamma_\nu}}{r_{A_{1g}}},\, {b_{\Gamma_\nu}^i}' = \frac{b_{\Gamma_\nu}}{b_{A_{1g}}}, \\ \nn
&{a_{\Gamma_\nu}^i}' = \frac{a_{\Gamma_\nu}}{\xi^2 (-r_{A_{1g}})},\, a_{T_{1u}\Gamma_\nu}' = \frac{a_{T_{1u}\Gamma_\nu}}{\xi (-r_{A_{1g}})}, \\ \nn
& \underset{\mu \neq T_{1u}}{a_{\Gamma_\mu \Gamma_\nu}'} = \frac{a_{\Gamma_\mu \Gamma_\nu}}{\xi^2 (-r_{A_{1g}})},\, {c_{A_{1g}\Gamma_\nu}^i}' = \frac{c_{A_{1g}\Gamma_\nu}}{b_{A_{1g}}},
\end{align}
and the covariant derivatives in the dimensionless units are
\begin{align}
D'_{i} = \partial_{i}' - i A_i',\ D'_{t} = \partial_{t}' + i \phi'.
\end{align}
In the following, writing the TDGL equations in dimensionless form, we omit the prime sign ($'$) in dimensionless quantities and the GL parameters. For a thin film case, for which we assume that dependence on the $z$ coordinate is very weak and derivatives with respect to the $z$ coordinate can be neglected, these equations take form
\begin{widetext}
\begin{align}
\label{eq: TDGL A1g}
\Gamma_{A_{1g}}\,D_t \eta_{A_{1g}}
&= r_{A_{1g}}\,\eta_{A_{1g}}
+ a_{A_{1g}}\sum_{i} D_i^{2}\eta_{A_{1g}}
- b_{A_{1g}}\lvert \eta_{A_{1g}}\rvert^{2}\eta_{A_{1g}}
+ 2\,a_{A_{1g},T_{1u}}\!\left(-\sum_{i} D_i\,\eta^{\,i}_{T_{1u}}\right) \\ \nn
&\quad
+ \tfrac12\,a_{A_{1g},E_g}\!\left(D_x^{2}\eta^{\,1}_{E_g}-D_x^{2}\eta^{\,2}_{E_g}
- D_y^{2}\eta^{\,1}_{E_g}-D_y^{2}\eta^{\,2}_{E_g}\right) \\ \nn
&\quad
+ a_{A_{1g},T_{1g}}\!\left(-D_xD_y\,\eta^{\,1}_{T_{1g}}+D_yD_x\,\eta^{\,1}_{T_{1g}}\right)
+ \alpha_{A_{1g},T_{2g}}\!\left(D_xD_y\,\eta^{\,1}_{T_{2g}}+D_yD_x\,\eta^{\,1}_{T_{2g}}\right) \\ \nn
&\quad -\sideset{}{'}\sum_{\Gamma \in \Gamma_r} c_{A_{1g}\Gamma}^1 \sum_{i=1}^{d_\Gamma} \abs{\eta_{\Gamma}^i}^2 \eta_{A_{1g}} - \sideset{}{'}\sum_{\Gamma \in \Gamma_r} c_{A_{1g}\Gamma}^2 \sum_{i=1}^{d_\Gamma} \left(\eta_{\Gamma}^i\right)^2 \eta_{A_{1g}}^*,
\end{align}
for the $s$-wave OP, where $r_{A_{1g}}=1$, $a_{A_{1g}}=1$, and $b_{A_{1g}}=1$,
\begin{align}
\label{eq: TDGL T1u}
\Gamma_{T_{1u}}\,D_t \eta^{\,1}_{T_{1u}}
&= r_{T_{1u}}\,\eta^{\,1}_{T_{1u}}
+ a^{1}_{T_{1u}}\,D_x^{2}\eta^{\,1}_{T_{1u}}
+ a^{2}_{T_{1u}}\,D_y^{2}\eta^{\,1}_{T_{1u}}
+ a^{3}_{T_{1u}}\,D_xD_y\,\eta^{\,2}_{T_{1u}}
+ a^{4}_{T_{1u}}\,D_yD_x\,\eta^{\,2}_{T_{1u}}
\\ \nn
&\quad
-2 b^{1}_{T_{1u}}\!\left(\sum_{i=1}^{3}\!\left|\eta^{\,i}_{T_{1u}}\right|^{2}\right)\eta^{\,1}_{T_{1u}}
-2 b^{2}_{T_{1u}}\!\left(\sum_{i=1}^{3}\big(\eta^{\,i}_{T_{1u}}\big)^2\right) \eta^{\,1*}_{T_{1u}}
- b^{3}_{T_{1u}}\!\left(\left|\eta^{\,2}_{T_{1u}}\right|^{2}+\left|\eta^{\,3}_{T_{1u}}\right|^{2}\right)\eta^{\,1}_{T_{1u}}
\\ \nn
&\quad
+ 2 a_{A_{1g}T_{1u}}\,D_x \eta_{A_{1g}}
+ a_{T_{1u}E_g}\!\left(D_x\eta^{\,1}_{E_g}-D_x\eta^{\,2}_{E_g}\right)
+ 2 a_{T_{1u}T_{1g}}\,D_y \eta^{\,1}_{T_{1g}}
- 2 a_{T_{1u}T_{2g}}\,D_y \eta^{\,1}_{T_{2g}}\\ \nn
&\quad  -c_{A_{1g}T_{1u}}^1 \abs{\eta_{A_{1g}}}^2 \eta_{T_{1u}}^1 - c_{A_{1g}T_{1u}}^2 \left(\eta_{A_{1g}}\right)^2 \eta_{T_{1u}}^{1*},
\end{align}
\begin{align}
\label{eq: TDGL T1u2}
\Gamma_{T_{1u}}\,D_t \eta^{\,2}_{T_{1u}}
&= r_{T_{1u}}\,\eta^{\,2}_{T_{1u}}
+ a^{1}_{T_{1u}}\,D_y^{2}\eta^{\,2}_{T_{1u}}
+ a^{2}_{T_{1u}}\,D_x^{2}\eta^{\,2}_{T_{1u}}
+ a^{3}_{T_{1u}}\,D_y D_x\,\eta^{\,1}_{T_{1u}}
+ a^{4}_{T_{1u}}\,D_x D_y\,\eta^{\,1}_{T_{1u}}
\\ \nn
&\quad
-2 b^{1}_{T_{1u}}\!\left(\sum_{i=1}^{3}\big|\eta^{\,i}_{T_{1u}}\big|^{2}\right)\eta^{\,2}_{T_{1u}}
-2 b^{2}_{T_{1u}}\!\left(\sum_{i=1}^{3}\big(\eta^{\,i}_{T_{1u}}\big)^{2}\right)\eta^{\,2*}_{T_{1u}}
- b^{3}_{T_{1u}}\!\left(\big|\eta^{\,1}_{T_{1u}}\big|^{2}+\big|\eta^{\,3}_{T_{1u}}\big|^{2}\right)^{2}\eta^{\,2}_{T_{1u}}
\\ \nn
&\quad
+ 2 a_{A_{1g}T_{1u}}\,D_y \eta_{A_{1g}}
+ a_{T_{1u}E_g}\!\left(-D_y\eta^{\,1}_{E_g}-D_y\eta^{\,2}_{E_g}\right)
- 2 a_{T_{1u}T_{1g}}\,D_x \eta^{\,1}_{T_{1g}}
- 2 a_{T_{1u}T_{2g}}\,D_x \eta^{\,1}_{T_{2g}}\\ \nn
&\quad -c_{A_{1g}T_{1u}}^1 \abs{\eta_{A_{1g}}}^2 \eta_{T_{1u}}^2 - c_{A_{1g}T_{1u}}^2 \left(\eta_{A_{1g}}\right)^2 \eta_{T_{1u}}^{2*},
\end{align}
\begin{align}
\label{eq: TDGL T1u3}
\Gamma_{T_{1u}}\,D_t \eta^{\,3}_{T_{1u}}
&= r_{T_{1u}}\,\eta^{\,3}_{T_{1u}}
+ a^{2}_{T_{1u}}\big(D_x^{2}+D_y^{2}\big)\eta^{\,3}_{T_{1u}}
\\ \nn
&\quad
-2 b^{1}_{T_{1u}}\!\left(\sum_{i=1}^{3}\lvert \eta^{\,i}_{T_{1u}}\rvert^{2}\right)\eta^{\,3}_{T_{1u}}
-2 b^{2}_{T_{1u}}\!\left(\sum_{i=1}^{3} \big( \eta^{\,i}_{T_{1u}}\big)^2 \right) \eta^{\,3*}_{T_{1u}}
- b^{3}_{T_{1u}}\!\left(\lvert \eta^{\,1}_{T_{1u}}\rvert^{2}+\lvert \eta^{\,2}_{T_{1u}}\rvert^{2}\right)\eta^{\,3}_{T_{1u}}
\\ \nn
&\quad
+ 2 a_{T_{1u}T_{1g}}\!\left(-D_x\eta^{\,2}_{T_{1g}}-D_y\eta^{\,3}_{T_{1g}}\right)
+ 2 a_{T_{1u}T_{2g}}\!\left(-D_x\eta^{\,2}_{T_{2g}}-D_y\eta^{\,3}_{T_{2g}}\right)\\ \nn
&\quad -c_{A_{1g}T_{1u}}^1 \abs{\eta_{A_{1g}}}^2 \eta_{T_{1u}}^3 - c_{A_{1g}T_{1u}}^2 \left(\eta_{A_{1g}}\right)^2 \eta_{T_{1u}}^{3*},
\end{align}
for the components of the $p$-wave OP, and
\begin{align}
\label{eq: TDGL Eg}
\Gamma_{E_g}\,D_t \eta^{\,1}_{E_g}
&= r_{E_g}\,\eta^{\,1}_{E_g}
+ a^{1}_{E_g}\sum_{i=1}^{3} D_i^{2}\eta^{\,1}_{E_g}
+ a^{2}_{E_g}\!\left(-D_x^{2}\eta^{\,1}_{E_g}-D_y^{2}\eta^{\,1}_{E_g}
+\sqrt{3}\,D_x^{2}\eta^{\,2}_{E_g}-\sqrt{3}\,D_y^{2}\eta^{\,2}_{E_g}\right) \\ \nn
&\quad
-2 b^{1}_{E_g}\!\left(\sum_{i=1}^{2}\lvert\eta^{\,i}_{E_g}\rvert^{2}\right)\eta^{\,1}_{E_g}
-2 b^{2}_{E_g}\!\left(\eta^{\,1*}_{E_g}\eta^{\,2}_{E_g}
-\eta^{\,2*}_{E_g}\eta^{\,1}_{E_g}\right)\eta^{\,2}_{E_g}   \\ \nn
&\quad
+\tfrac12\,a_{A_{1g}E_g}\!\left(D_x^{2}\eta_{A_{1g}}-D_y^{2}\eta_{A_{1g}}\right)
- a_{T_{1u}E_g}\!\left(D_x\eta^{\,1}_{T_{1u}}-D_y\eta^{\,2}_{T_{1u}}\right) \\ \nn
&\quad
+ a^{1}_{E_gT_{1g}}\!\left(-\tfrac{1}{3}D_xD_y\eta^{\,1}_{T_{1g}}
-\tfrac{1}{3}D_yD_x\eta^{\,1}_{T_{1g}}\right)
+ a^{2}_{E_gT_{1g}}\!\left(-\tfrac{1}{3}D_xD_y\eta^{\,1}_{T_{1g}}
-\tfrac{1}{3}D_yD_x\eta^{\,1}_{T_{1g}}\right) \\ \nn
&\quad
+ a^{1}_{E_gT_{2g}}\!\left(\tfrac{1}{3}D_xD_y\eta^{\,1}_{T_{2g}}
-\tfrac{1}{3}D_yD_x\eta^{\,1}_{T_{2g}}\right)
+ a^{2}_{E_gT_{2g}}\!\left(-\tfrac{1}{3}D_xD_y\eta^{\,1}_{T_{2g}}
+\tfrac{1}{3}D_yD_x\eta^{\,1}_{T_{2g}}\right)\\ \nn
&\quad -c_{A_{1g}T_{1u}}^1 \abs{\eta_{A_{1g}}}^2 \eta_{E_g}^1 - c_{A_{1g}T_{1u}}^2 \left(\eta_{A_{1g}}\right)^2 \eta_{E_g}^{1*},
\end{align}
\begin{align}
\label{eq: TDGL Eg2}
\Gamma_{E_g}\,D_t \eta^{\,2}_{E_g}
&= r_{E_g}\,\eta^{\,2}_{E_g}
+ a^{1}_{E_g}\sum_{i=1}^{3} D_i^{2}\eta^{\,2}_{E_g}
+ a^{2}_{E_g}\!\left( D_x^{2}\eta^{\,2}_{E_g}+D_y^{2}\eta^{\,2}_{E_g}
+ \sqrt{3}\,D_x^{2}\eta^{\,1}_{E_g}-\sqrt{3}\,D_y^{2}\eta^{\,1}_{E_g}\right) \\ \nn
&\quad
- 2 b^{1}_{E_g}\!\left(\sum_{i=1}^{2}\lvert\eta^{\,i}_{E_g}\rvert^{2}\right)\eta^{\,2}_{E_g}
+ 2 b^{2}_{E_g}\!\left(\eta^{\,1*}_{E_g}\eta^{\,2}_{E_g}
-\eta^{\,2*}_{E_g}\eta^{\,1}_{E_g}\right)\eta^{\,1}_{E_g} \\ \nn
&\quad
+\tfrac12\,a_{A_{1g}E_g}\!\left(-D_x^{2}\eta_{A_{1g}}-D_y^{2}\eta_{A_{1g}}\right)
- a_{T_{1u}E_g}\!\left(-D_x\eta^{\,1}_{T_{1u}}-D_y\eta^{\,2}_{T_{1u}}\right) \\ \nn
&\quad
+ a^{1}_{E_gT_{1g}}\!\left(D_xD_y\,\eta^{\,1}_{T_{1g}}-D_yD_x\,\eta^{\,1}_{T_{1g}}\right)
+ a^{2}_{E_gT_{1g}}\!\left(-D_xD_y\,\eta^{\,1}_{T_{1g}}+D_yD_x\,\eta^{\,1}_{T_{1g}}\right) \\ \nn
&\quad
+ a^{1}_{E_gT_{2g}}\!\left(-D_xD_y\,\eta^{\,1}_{T_{2g}}
-D_yD_x\,\eta^{\,1}_{T_{2g}}\right)
+ a^{2}_{E_gT_{2g}}\!\left(-D_xD_y\,\eta^{\,1}_{T_{2g}}
-D_yD_x\,\eta^{\,1}_{T_{2g}}\right)\\ \nn
&\quad -c_{A_{1g}T_{1u}}^1 \abs{\eta_{A_{1g}}}^2 \eta_{E_g}^2 - c_{A_{1g}T_{1u}}^2 \left(\eta_{A_{1g}}\right)^2 \eta_{E_g}^{2*},
\end{align}
for the $E_g$ ($d_{x^2-y^2}$ and $d_{z^2}$) components of the $d$-wave OP. We write down the rest of the TDGL equations in~\cref{app:restofthetdlgeqs}.
\end{widetext}

\subsection{Current}
The variation of~\cref{eq: full static Lagrangian} with respect to the vector potential yields the expression for the superconducting current. We compactly write terms in the Lagrangian density corresponding to the singlet-to-singlet and triplet-to-triplet couplings as
\begin{align}
\label{eq: Q parts}
\mathcal{L}_{\mu \nu} = Q_{\mu \nu, ij}^{\alpha \beta} \left(D_i \eta_\mu^\alpha\right)^* D_j \eta_\nu^\beta,\ \mathrm{with}\  Q_{\mu \nu, ij}^{\alpha \beta} = Q_{\nu \mu, ji}^{\beta \alpha}
\end{align}
and to the triplet-to-singlet couplings as
\begin{align}
\label{eq: P parts}
\mathcal{L}_{T_{1u} \nu} = P_{\nu, i}^{\alpha \beta} \left( \left(D_i \eta_{T_{1u}}^\alpha\right)^* \eta_\nu^\beta - \eta_{T_{1u}}^{\alpha*} D_i\eta_\nu^\beta +c.c \right).
\end{align}
The values for $Q_{\mu \nu, ij}^{\alpha \beta}$ and $P_{\nu, i}^{\alpha \beta}$ are given in~\cref{app:coefficients}. For the SC current density, we find
\begin{align}
\label{eq: SC current}
\b{j_s}_i = \frac{4e}{\hbar} \Im{Q_{\mu \nu, ij}^{\alpha \beta} \eta_\mu^{\alpha*} D_j \eta_\nu^\beta} - \frac{4e}{\hbar} \Im{P_{\nu, i}^{\beta \alpha} \eta_\nu^{\alpha*} \eta_{T_{1u}}^\beta}.
\end{align}
To find the full current, we add the normal component $\b{j_n}$
\begin{align}
\label{eq: full current}
\b{j} &=\b{j_s}+\b{j_n},\ \mathrm{where} \\ \nn
\b{j_n} &= \sigma \b{E} = -\sigma \left(\grad \phi + \frac{\partial \b{A}}{\partial t} \right).
\end{align}
The electric potential $\phi$ is then found from the continuity equation
\begin{align}
\grad \cdot \b{j} = 0,
\end{align}
in which the charge-neutrality condition $\partial \rho/\partial t$, where $\rho$ is the charge density, is assumed. In the dimensionless (primed) units, this yields
\begin{align}
\label{eq: electric potential eq.}
{\laplacian}' \phi' = \grad' \cdot \b{j_s}' - \frac{\partial (\grad' \cdot \b{A}')}{\partial t'}. 
\end{align}
In the dimensionless units, the expression for current is the same as in~\cref{eq: SC current}, except for the absence of the $4e/\hbar$ coefficients, and the expressions for dimensionless $Q_{\mu \nu, ij}^{\alpha \beta'}$ are the same as $Q_{\mu \nu, ij}^{\alpha \beta}$ with the dimensional parameters being substituted with the corresponding dimensionless (primed) ones.

\subsection{Boundary conditions}
The system of the TDGL~\cref{eq: TDGL A1g,eq: TDGL T1u,eq: TDGL T1u2,eq: TDGL T1u3,eq: TDGL Eg,eq: TDGL Eg2,eq: TDGL T1g,eq: TDGL T1g2,eq: TDGL T1g3,eq: TDGL T2g,eq: TDGL T2g2,eq: TDGL T2g3} and~\cref{eq: electric potential eq.} has to be augmented with appropriate boundary conditions, which can be found by requiring the surface term, arising as the result of the variation of the free energy with respect to $\eta_\nu^*$, to be equal to zero. Generally, depending on the type of the material at the interface with the SC, there is also a surface term contribution to the Free energy~\cite{samoilenka2021}, i.e., $F=\int d^d x \mathcal{L}_{st} + F_{surf}$. We assume $F_{surf}=0$ here, which is the case for a boundary with an insulator~\cite{samoilenka2021,deGennes2018,caroli1962,abrikosov1965}. Then the boundary conditions take form
\begin{align}
n_i \cdot \left( Q_{\mu \nu, ij}^{\alpha \beta} D_j \eta_\nu^\beta - P_{\nu, i}^{\beta \alpha} \eta_{T_{1u}}^\beta \right) = 0, \\ \nn
n_i \cdot P_{\nu, i}^{\alpha \beta} \eta_{\nu}^\beta = 0,
\end{align}
where $n_i$ are the components of the normal to the interface unit vector $\b{n}$. Note that these two equations lead to that that the normal to the interface components of both the singlet and triplet components of the SC current vanish at the interface.

For numerical simplicity, in the case of a finite-width Gaussian beam, we employ simpler boundary conditions. Expecting that outside the effective area of the beam spot, the induced triplet and singlet components are nearly zero, we use zero boundary conditions for those components:
\begin{align}
\eta_{T1u}^i = 0,\ i=1..3, \\ \nn
\eta_{Eg}^i = 0,\ i=1,2, \\ \nn
\eta_{T1g}^i = 0,\ i=1..3, \\ \nn
\eta_{T2g}^i = 0,\ i=1..3, \\
\end{align}
and the usual boundary condition for the $s$-wave component of the SC current
\begin{align}
\b{n} \cdot \b{D} \eta_{A1g} = 0.
\end{align}
This 
simplification agrees with the results of simulations.

The boundary conditions for~\cref{eq: electric potential eq.} are 
\begin{align}
    \b{n}\cdot \left( \grad \phi + \frac{\partial \b{A}}{\partial t} \right) = 0,
\end{align}
which corresponds to zero normal current flowing through the interfaces.

\section{Results for Infinitely wide uniform and Gaussian beams}
\label{sec:results}

\begin{figure*}[!htbp]
    \centering
    \begin{subfigure}[t]{0.45\textwidth}
    \centering
    \includegraphics[width=\linewidth]{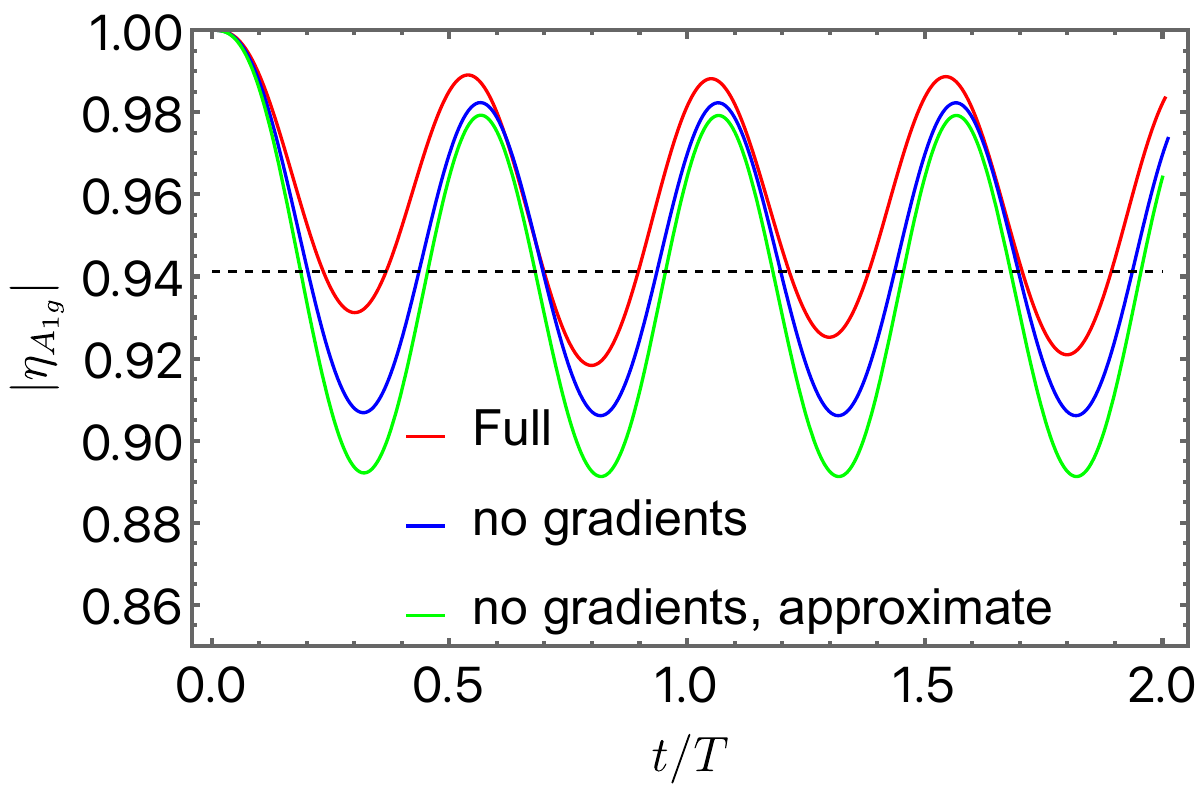}
    \caption{}
    \end{subfigure}
    \begin{subfigure}[t]{0.45\textwidth}
    \centering
    \includegraphics[width=\textwidth]{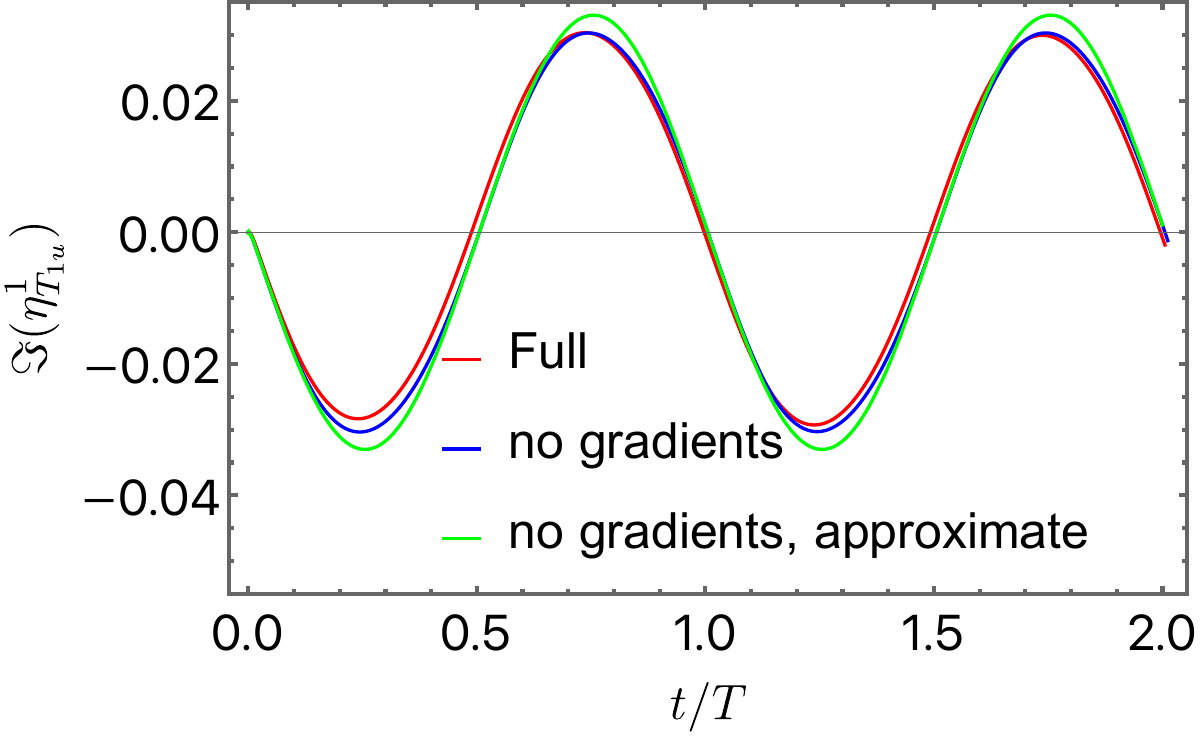}
    \caption{}
    \end{subfigure}
    \begin{subfigure}[t]{0.45\textwidth}
    \centering
    \includegraphics[width=\textwidth]{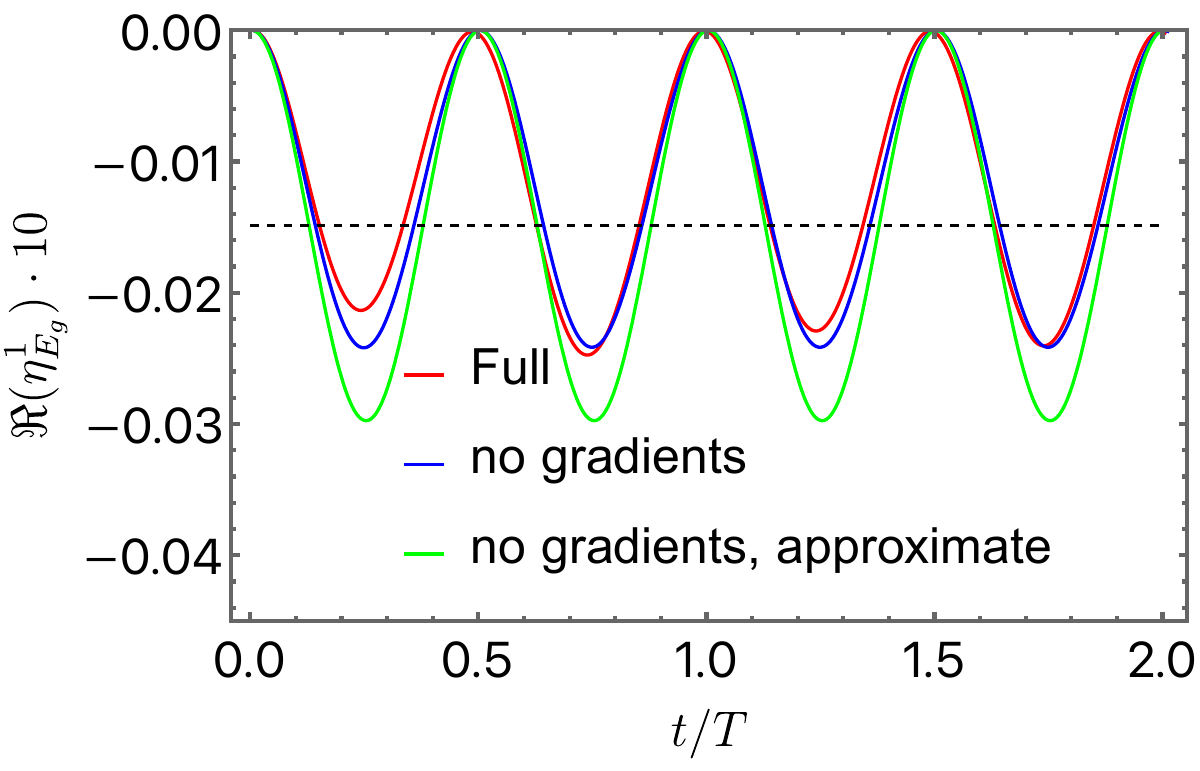}
    \caption{}
    \end{subfigure}
    \begin{subfigure}[t]{0.45\textwidth}
    \centering
    \includegraphics[width=\linewidth]{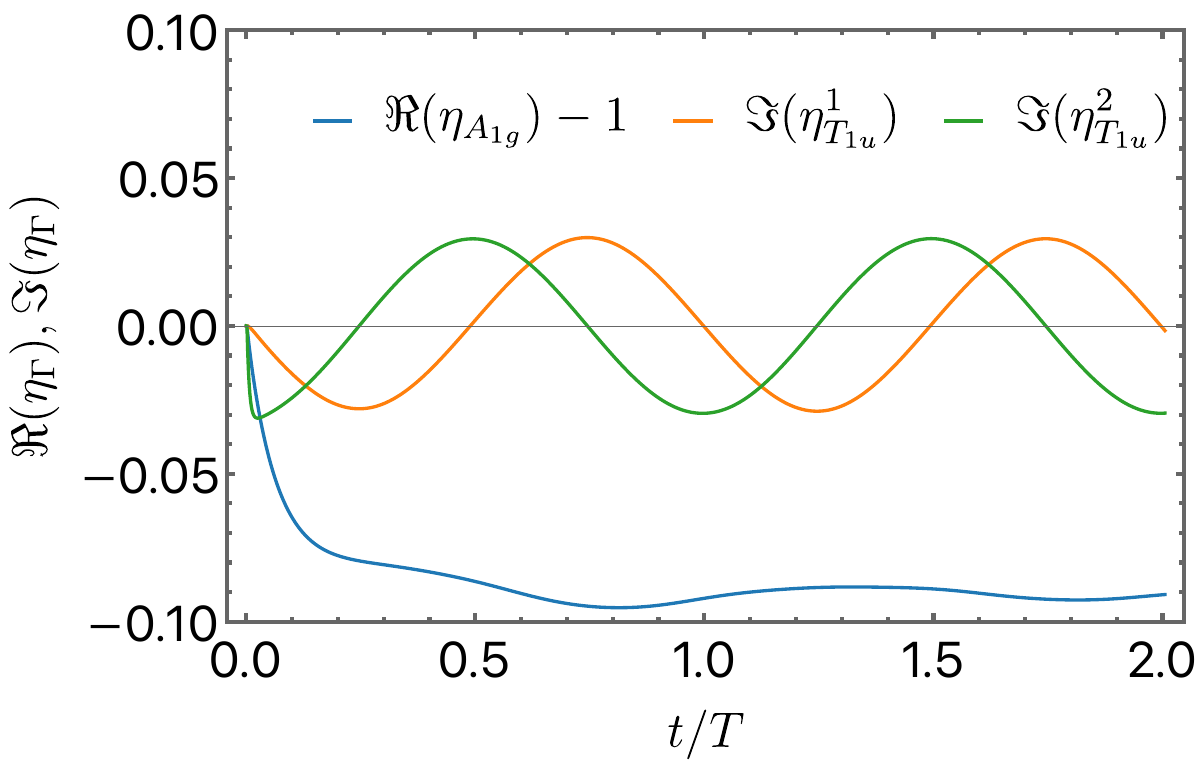}
    \caption{}
    \end{subfigure}
    \caption{(a) $\abs{\eta_{A_{1g}}}$ ($s$-wave); (b) $\Im(\eta_{T_{1u}}^1)$ ($p$-wave) ; (c) $\Re(\eta_{E_{g}}^1)$ ($d$-wave). Red and blue curves correspond to the solutions of~\cref{eq: TDGL A1g,eq: TDGL T1u,eq: TDGL T1u2,eq: TDGL T1u3,eq: TDGL Eg,eq: TDGL Eg2,eq: TDGL T1g,eq: TDGL T1g2,eq: TDGL T1g3,eq: TDGL T2g,eq: TDGL T2g2,eq: TDGL T2g3} in the case of a Gaussian and uniform linearly polarized beams, respectively. Black curve corresponds to the approximate solution of~\cref{eq: simplified eqs.,eq: simplified eqs.2,eq: simplified eqs.3,eq: simplified eqs.4,eq: simplified eqs.5,eq: simplified eqs.6} in the case of a uniform beam. Dashed lines correspond to the approximate average values given by~\cref{eq: average values} around which the approximate solutions oscillate. The TDGL parameters are given in~\cref{tab: Coefficients}, the dimensionless frequency and amplitude of the microwaves are $\omega=1$ and $A_{0}=0.5$, respectively, the waist of the Gaussian beam is $w_0=4 \xi$, and the side length of the thin square film is $l=20\xi$.    (d) Time-dependence of the values of $\eta_{A_{1g}}$ ($s$-wave), $\eta_{T_{1u}}^1$ ($p_y$-wave in the 2D limit), and $\eta_{T_{1u}}^2$ ($p_x$-wave in the 2D limit) order parameters at the center of the circularly polarized Gaussian beam with waist $w_0=4 \xi$. The TDGL parameters are given in~\cref{tab: Coefficients}, the dimensionless frequency and amplitude of the beam are $\omega=1$ and $A_{0}=0.5$, respectively, the waist of the beam is $w_0=4 \xi$, and the side length of the thin square film is $l=20\xi$.} 
\label{fig: Plots of OPs vs t}
\end{figure*}

We start with considering a uniform infinitely wide beam. In that case, a natural assumption is that the solution also has to be uniform. Thus, we neglect the spatial derivatives in the TDGL~\cref{eq: TDGL A1g,eq: TDGL T1u,eq: TDGL T1u2,eq: TDGL T1u3,eq: TDGL Eg,eq: TDGL Eg2,eq: TDGL T1g,eq: TDGL T1g2,eq: TDGL T1g3,eq: TDGL T2g,eq: TDGL T2g2,eq: TDGL T2g3}. The numerical solution of the resulting system of ODEs for the values of the SC OPs with TDGL parameters given in~\cref{tab: Coefficients} and the incident $\hat{x}$-polarized microwave with dimensionless frequency $\omega=1$ and amplitude $A_{0}=0.5$ is plotted in~\cref{fig: Plots of OPs vs t} in red curves. Microscopic validity of the $s$-wave TDGL eqs. assumes that $\Gamma_{A_{1g}} \ll \tau_E^{-1}$~\cite{watts-tobin1981,kopnin2001theory}, where $\tau_E$ is the inelastic electron-phonon scattering time. For clean systems, $\tau_E^{-1} \sim \frac{k_B T^3}{\hbar \Omega_D^2}$, where $\Omega_D$ is the Debye temperature. For low-temperature superconductors such as Nb, $\tau_E^{-1} \sim 10^9$ s$^{-1}$; for superconductors like Nb$_3$Al and MgB$_2$, $\tau_E^{-1}$ could be $\sim 10^{10}$ s$^{-1}$; and in cuprates $\tau_E^{-1}$ could be $\sim 10^{11}$ s$^{-1}$. In dirty superconductors, $\tau_E^{-1}$ might be larger~\cite{sergeev2000}. For example, for dirty NbN thin films $\tau_E \sim 10^{11}$ s$^{-1}$ has been reported~\cite{sidorova2020}.   
Thus, the relevant frequencies of the incident EM radiation is in microwave range.  
In Ref.~\cite{zhu1998}, a time-dependent GL theory for a mixed $s$- and $d$-wave SC with spherically symmetric Fermis surface was derived, which showed that the corresponding gradient and quartic intra-component GL parameters and inter-component GL parameters are on the same order. The same is true for the GL theory with mixed $s$- and $d$-wave OPs derived from a microscopic model with nearest-neighbor pairing interactions on a square lattice~\cite{talkachov2025}. We assume here that this is true for the rest of the SC components. In this regard, we note that the microscopically derived value in Ref.~\cite{gassner2024} of the coefficient for the $s$-to-$p$ coupling, $a_{A_{1g}T_{1u}}$, in a honeycomb lattice with Kane-Mele spin-orbit coupling model is on the same order as the coefficients for the $s$-to-$s$, $a_{A_{1g}}$, and $p$-to-$p$, $a_{T1u}^1$, gradient terms in $\mathcal{L}_{A_{1g}}$ and $\mathcal{L}_{T_{1u}}$, respectively~\cite{gassnerpersonal}. We also set the inertial coefficients for each SC component to be the same, i.e., $\Gamma_{\nu}=\Gamma$.

\begin{center}
\begin{table*}[!htbp]
\begin{tabular}{|c|c| }
\hline
 $A_{1g}$ & $\Gamma_{A_{1g}}=1$, $r_{A_{1g}}=1,b_{A_{1g}} = 1, a_{A_{1g}} = 1, a_{A_{1g}T_{1u}} = 1, a_{A_{1g}E_g} = 1, a_{A_{1g}T_{1g}} = 1, a_{A_{1g}T_{2g}} = 1$  \\ 
 \hline
 $T_{1u}$ & $\Gamma_{T_{1u}}=1, r_{T_{1u}} = -30; a_{T_{1u}}^1 = 1.1, a_{T_{1u}}^2 = 1; a_{T_{1u}}^3 = 1.2, a_{T_{1u}}^4 = 1.1, b_{T_{1u}}^1 = 1, b_{T_{1u}}^2 = 1, b_{T_{1u}}^3 = 1, a_{T_{1u}E_{g}} = 1, a_{T_{1u}T_{1g}} = 1.2,$ \\
 & $a_{T_{1u}T_{2g}} = 1,
a_{T_{1u}T_{1g}} = 1.2, a_{T_{1u}T_{2g}} = 1$, $c_{A_{1g}T_{1u}}^1=c_{A_{1g}T_{1u}}^2=1$  \\  
 \hline
 $E_{g}$ & $\Gamma_{E_{g}}=1, r_{E_g} = -40, a_{E_g}^1 = 1.05, a_{E_g}^2 = 1, b_{E_g}^1 = 1, b_{E_g}^2 = 1, a_{E_gT_{1g}}^1 = 1.1, a_{E_gT_{1g}}^2 = 1, a_{E_gT_{2g}}^1 = 1.1, a_{E_gT_{2g}}^2 = 1.05$\\
  & $c_{A_{1g}E_{g}}^1=c_{A_{1g}E_{g}}^2=1$ \\
 \hline
 $T_{1g}$ & $\Gamma_{T_{1g}}=1, r_{T_{1g}} = -50, a_{T_{1g}}^1 = 1, a_{T_{1g}}^2 = 1.05, a_{T_{1g}}^3 = 1.1, a_{T_{1g}}^4 = 0.95, b_{T_{1g}}^1 = 1, b_{T_{1g}}^2 = 1, b_{T_{1g}}^3 = 1,
a_{T_{1g}T_{2g}}^1 = 1, a_{T_{1g}T_{2g}}^2 = 1.1,$\\
 & $ a_{T_{1g}T_{2g}}^3 = 1.2$, $c_{A_{1g}T_{1g}}^1=c_{A_{1g}T_{1g}}^2=1$ \\
 \hline
 $T_{2g}$ & $\Gamma_{T_{2g}}=1, r_{T_{2g}} = -60, a_{T_{2g}}^1 = 0.9, a_{T_{2g}}^2 = 0.95, a_{T_{2g}}^3 = 0.85, aT_{T_{2g}}^4 = 0.8, b_{T_{2g}}^1 = 1, b_{T_{2g}}^2 = 1, b_{T_{2g}}^3 = 1$, \\
  & $c_{A_{1g}T_{2g}}^1=c_{A_{1g}T_{2g}}^2=1$\\
\hline
\end{tabular}
\caption{Coefficients in the TDGL model used for the simulations that produced data for~\cref{fig: Plots of OPs vs t}.}
\label{tab: Coefficients}
\end{table*}
\end{center}

To gain analytical insight, under the assumption that the change in $\eta_{A_{1g}}$, $\delta \eta_{A_{1g}} = 1 - \eta_{A_{1g}}$, and the induced components are small. We solve approximate decoupled equations, where $1$s now stand for the unperturbed $\eta_{A_{1g}}$:
\begin{widetext}
\begin{align}
\label{eq: simplified eqs.}
\Gamma_{A_{1g}} d_t {\eta}_{A_{1g}}
&=\Bigl(1-\eta_{A_{1g}}^2\Bigr)\,\eta_{A_{1g}}
 - A_x^{2}\,\eta_{A_{1g}},
\\[4pt]
\label{eq: simplified eqs.2}
\Gamma_{T_{1u}} d_t  {\eta}_{T_{1u}}^1
&= r_{T_{1u}}\,\eta_{T_{1u}}^1
 - a_{T_{1u}}^1\,A_x^{2}\,\eta_{T_{1u}}^1
 - 2\,a_{A_{1g}T_{1u}}\,i\,A_x\cdot 1
 - c_{A_{1g}T_{1u}}^1\,\eta_{T_{1u}}^1
 + c_{A_{1g}T_{1u}}^2\,\eta_{T_{1u}}^{1*},
\\[4pt]
\label{eq: simplified eqs.3}
\Gamma_{T_{1u}} d_t {\eta}_{T_{1u}}^2
&= r_{T_{1u}}\,\eta_{T_{1u}}^2
 - a_{T_{1u}}^2\,A_x^{2}\,\eta_{T_{1u}}^2
 - c_{A_{1g}T_{1u}}^1\,\eta_{T_{1u}}^2
 - c_{A_{1g}T_{1u}}^2\,\eta_{T_{1u}}^{2*},
\\[4pt]
\label{eq: simplified eqs.4}
\Gamma_{T_{1u}} d_t {\eta}_{T_{1u}}^3
&= r_{T_{1u}}\,\eta_{T_{1u}}^3
 - a_{T_{1u}}^2\,A_x^{2}\,\eta_{T_{1u}}^3
 - c_{A_{1g}T_{1u}}^1\,\eta_{T_{1u}}^3
 - c_{A_{1g}T_{1u}}^2\,\eta_{T_{1u}}^{3*},
\\[4pt]
\label{eq: simplified eqs.5}
\Gamma_{E_g} d_t {\eta}_{E_g}^1
&= r_{E_g}\,\eta_{E_g}^1
 - \tfrac{1}{2}\,a_{A_{1g}E_g}\,A_x^{2}\cdot 1
 - c_{A_{1g}E_g}^1\,\eta_{E_g}^1
 - c_{A_{1g}E_g}^2\,\eta_{E_g}^{1*},
\\[4pt]
\label{eq: simplified eqs.6}
\Gamma_{E_g} d_t {\eta}_{E_g}^2
&= r_{E_g}\,\eta_{E_g}^2
 + \tfrac{1}{2}\,a_{A_{1g}E_g}\,A_x^{2}\cdot 1
 - c_{A_{1g}E_g}^1\,\eta_{E_g}^2
 - c_{A_{1g}E_g}^2\,\eta_{E_g}^{2*},
\end{align}
\end{widetext}
omitting the ones for $\eta_{T_{1g}}^i$ and $\eta_{T_{2g}}^i$ as these components remain essentially zero for the chosen parameters.

The solutions to~\cref{eq: simplified eqs.,eq: simplified eqs.2,eq: simplified eqs.3,eq: simplified eqs.4,eq: simplified eqs.5,eq: simplified eqs.6}, linearized in $\delta \eta_{A_{1g}}$, are 
\begin{widetext}
\begin{align}
\eta_{A_{1g}}(t)
&= 1-\,\frac{A_0^2}{\Gamma_{A_{1g}}}\,
e^{
 -\frac{2t}{\Gamma_{A_{1g}}}
 - \frac{A_0^{2}\,t}{2\,\Gamma_{A_{1g}}}
 + \frac{A_0^{2}\,\sin(2\omega t)}{4\,\omega\,\Gamma_{A_{1g}}}
}
\int_{0}^{t}
\sin^2(\omega t_1)\,
e^{
 \frac{2t_1}{\Gamma_{A_{1g}}}
 + \frac{A_0^{2}\,t_1}{2\,\Gamma_{A_{1g}}}
 - \frac{A_0^{2}\,\sin(2\omega t_1)}{4\,\omega\,\Gamma_{A_{1g}}}
}\,
dt_1,\\ \nn
\eta_{T_{1u}}^1(t)
&=-i
e^{
 \Gamma_{T_{1u}}^{-1}\!\left(
   \tilde r_{T_{1u}}\,t
   - \frac{a_{T_{1u}} A_0^{2}}{2}\,t
   - \frac{a_{T_{1u}} A_0^{2}}{4}\,\sin(2\omega t)
 \right)}
\int_{0}^{t}
  2\,a_{A_{1g}T_{1u}}\,A_0 \sin(\omega t_1)\,
  e^{
   -\Gamma_{T_{1u}}^{-1}\!\left(
     \tilde r_{T_{1u}}\,t_1
     - \frac{a_{T_{1u}} A_0^{2}}{2}\,t_1
     - \frac{a_{T_{1u}} A_0^{2}}{4}\,\sin(2\omega t_1)
   \right)}
\,dt_1, \\ \nn
\eta_{T_{1u}}^2(t) &= \eta_{T_{1u}}^3(t) = 0, \\ \nn
\eta_{E_{g}}^1(t) &= -\eta_{E_{g}}^2(t)
= -\,\frac{a_{A_{1g}E_g}\,A_0^{2}}{4}\,
\frac{
 4\omega^{2}\Gamma_{E_g}^{2}\,e^{\tilde{r}_{E_g}/\Gamma_{E_g} t}
 - 4\omega^{2}\Gamma_{E_g}^{2} - \tilde{r}_{E_g}^2
 + \tilde{r}_{E_g}^{2}\cos(2\omega t)
 - 2\,\tilde{r}_{E_g}\,\omega\,\Gamma_{E_g}\sin(2\omega t)
}{
 \tilde{r}_{E_g}^{3} + 4 \tilde{r}_{E_g} \omega^{2}\Gamma_{E_g}^{2}
}\,, \\ \nn
\eta_{T_{1g}}^1(t) &= \eta_{T_{1g}}^2(t)= \eta_{T_{1g}}^3(t) = 0,\, \eta_{T_{2g}}^1(t) = \eta_{T_{2g}}^2(t)= \eta_{T_{2g}}^3(t) = 0,
\end{align}
\end{widetext}
where $\tilde{r}_{T_{1u}}=r_{T_{1u}} - (c_{A_{1g}T_{1u}}-c_{A_{1g}T_{1u}})$ and $\tilde{r}_{E_{g}}=r_{E_{g}} - (c_{A_{1g}E_{g}}+c_{A_{1g}E_{g}})$.

When $\abs{\frac{A_0^{2}\,t}{2\,\Gamma_{A_{1g}}}
 - \frac{A_0^{2}\,\sin(2\omega t)}{4\,\omega\,\Gamma_{A_{1g}}}} \ll \abs{\frac{2t}{\Gamma_{A_{1g}}}}$ and $\abs{\frac{a_{T_{1u}} A_0^{2}}{2}\,t
   - \frac{a_{T_{1u}} A_0^{2}}{4}\,\sin(2\omega t)} \ll \abs{\tilde r_{T_{1u}}\,t}$, then

\begin{widetext}
\begin{align}
\eta_{A_{1g}}(t)
&= 1-\,\frac{A_0^2}{4(1+\omega^2 \Gamma_{A_{1g}}^2)}\,
\left(
 \omega^2 \Gamma_{A_{1g}}^2 e^{-2t/\Gamma_{A_{1g}}} - (1+\omega^2 \Gamma_{A_{1g}}^2) + \cos(2 \omega t) + \omega \Gamma_{A_{1g}} \sin(2 \omega t) \right),\\ \nn
\eta_{T_{1u}}^1(t)
&= -i\frac{2 A_0 a_{A_{1g}T_{1u}}}{\tilde{r}_{T_{1u}}^2+\omega^2 \Gamma_{T_{1u}}^2}\,
\left(
 \omega \Gamma_{T_{1u}} e^{-\tilde{r}_{T_{1u}} t/\Gamma_{T_{1u}}} + \omega \Gamma_{T_{1u}} \cos(\omega t) + \tilde{r}_{T_{1u}} \sin(\omega t) \right),
\end{align}
\end{widetext}
respectively.

The solutions have a form $\eta_\Gamma=f_{\Gamma}(t)g_{\Gamma}$, where $f_{\Gamma} \rightarrow 1$ as $t$ increases and $g_{\Gamma}$ is a periodic function with period $T=\pi/\omega$. Averaging over the period $T$ in the region where $f(g)\approx1$, from~\cref{eq: simplified eqs.,eq: simplified eqs.2,eq: simplified eqs.3,eq: simplified eqs.4,eq: simplified eqs.5,eq: simplified eqs.6}, we find values of $\eta_\Gamma$ about which oscillations occur
\begin{align}
\label{eq: average values}
    \eta_{A_{1g}0} &= 1 - \frac{A_0^2}{4+A_0^2}, \\ \nn
    \eta_{T_{1u}0} &= 0, \\ \nn
    \eta_{E_{g}0}^1 &=-\eta_{E_{g}0}^2=\frac{a_{A_{1g}E_g}}{\tilde{r}_{E_g}}A_0^2.
\end{align}
One can see that the induced singlet components oscillate around non-zero values due to rectification.
The amplitudes of the oscillations $\delta \eta_\Gamma$ can be found approximately as
\begin{align}
\delta \eta_{A_{1g}0} &= \frac{A_0^2}{4(1+\omega^2 \Gamma_{A_{1g}}^2)} \left( \cos(\phi_1) + \omega \Gamma_{A_{1g}}\sin(\phi_1) \right), \\
\delta \eta_{T_{1u}0}^1 &= i \frac{2 A_0 a_{A_{1g}T_{1u}}}{\tilde{r}_{T_{1u}}^2+\omega^2 \Gamma_{T_{1u}}^2} \left( \omega \Gamma_{T_{1u}} \cos(\phi_2) + \tilde{r}_{T_{1u}} \sin(\phi_2)) \right), \\
\delta \eta_{E_{g}0}^{1,2} &= -\frac{A_0^2 a_{A_{1g}E_{g}}}{4(\tilde{r}_{E_{g}}^2+4\omega^2 \Gamma_{E_{g}}^2)} \left( \tilde{r}_{E_{g}} \cos(\phi_3) -2 \omega \Gamma_{E_{g}}\sin(\phi_3) \right),
\end{align}
where $\phi_1=\arctan (\omega \Gamma_{A_{1g}})$, $\phi_2=\arctan (\frac{\tilde{r}_{T_{1u}}}{\omega \Gamma_{T_{1u}}})$, $\phi_3=\arctan (\frac{2\omega \Gamma_{E_{g}}}{\tilde{r}_{E_{g}}})$.

When $\frac{2\omega \Gamma_{T_{1u}}}{\tilde{r}_{T_{1u}}}\ll1$ and $\frac{2\omega \Gamma_{E_{g}}}{\tilde{r}_{E_{g}}}\ll1$, 
\begin{align*}
\delta \eta_{T_{1u}0}^1 &\approx i \frac{2 A_0 a_{A_{1g}T_{1u}} \tilde{r}_{T_{1u}}}{\tilde{r}_{T_{1u}}^2+\omega^2 \Gamma_{T_{1u}}^2},\ \mathrm{and} \\ \nn
\delta \eta_{E_{g}0}^{1,2} &\approx \frac{A_0^2 a_{A_{1g}E_{g}} \tilde{r}_{E_g}}{4(\tilde{r}_{E_{g}}^2+4\omega^2 \Gamma_{E_{g}}^2)},
\end{align*}
respectively.

To assess the effect of spatial derivatives, we simulate the dynamics of a superconducting condensate in a thin square film with a side length $l=20\xi$ subjected to a linearly polarized Gaussian beam with frequency $\omega/\Gamma^{-1}=1$ and waist $w_0=4 \xi$ with a center of the beam incident at the center of the film. As we are not interested in the exact dynamics, we also neglect the change in the scalar potential $\phi$. The corresponding vector potential is given by
\begin{align}
    \b{A}(\b{r},t) = A_0 e^{-\frac{r^2}{w_0^2}} \sin (\omega t) \hat{\b{x}}.
\end{align}

In~\cref{fig: Plots of OPs vs t}(a)--(c), we plot the absolute values of $\eta_{A_{1g}}$, the imaginary part of $\eta_{T_{1u}}^1$, and the real part of $\eta_{E_g}^1 \approx -\eta_{E_g}^2$ in the center of the film, the corresponding solutions of~\cref{eq: TDGL A1g,eq: TDGL T1u,eq: TDGL T1u2,eq: TDGL T1u3,eq: TDGL Eg,eq: TDGL Eg2,eq: TDGL T1g,eq: TDGL T1g2,eq: TDGL T1g3,eq: TDGL T2g,eq: TDGL T2g2,eq: TDGL T2g3} for the uniform beam, and the solutions of the approximate~\cref{eq: simplified eqs.,eq: simplified eqs.2,eq: simplified eqs.3,eq: simplified eqs.4,eq: simplified eqs.5,eq: simplified eqs.6}. The imaginary part of $\eta_{A_{1g}}$, the real part of $\eta_{T_{1u}}^1$, and the imaginary part of $\eta_{E_g}^1 \approx -\eta_{E_g}^2$ in the center of the film remain almost zero. The absolute value of $\eta_{T_{1u}}^2$ reaches an appreciable non-zero value away from the center of the film, but still about 20 times smaller than the maximum value of $\eta_{T_{1u}}^1$. The density plots that illustrate this are provided in~\cref{app:densityplots}.

In case of a linearly polarized beam, the average value of the induced triplet component is zero. In order to have non-zero value of the induced triplet component, a circularly polarized microwave beam can be utilized; see~\cref{fig: Plots of OPs vs t}(d) and~\cref{fig: Plots of OPs vs t circular1,fig: Plots of OPs vs t circular2,fig: Plots of OPs vs t circular3,fig: Plots of OPs vs t circular4}.

\begin{figure*}[!htbp]
    \centering
    \begin{subfigure}[t]{0.95\textwidth}
    \centering
    \includegraphics[width=\linewidth]{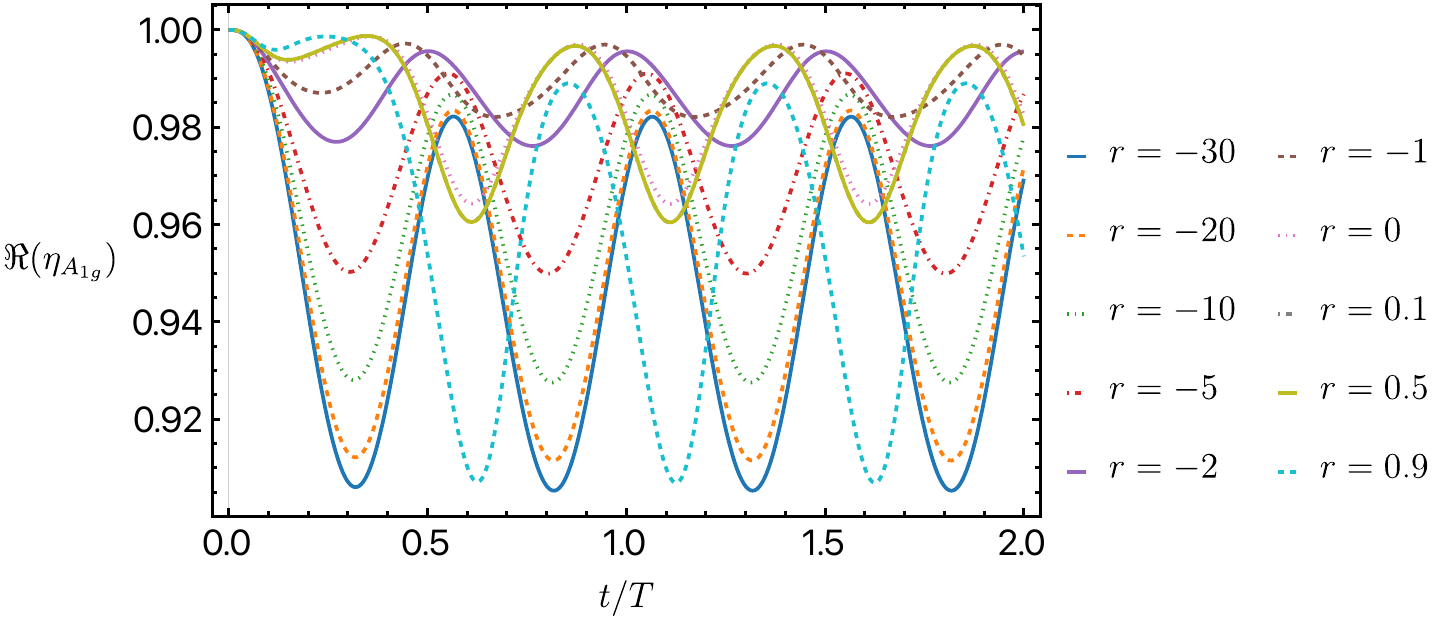}
    \caption{}
    \end{subfigure}
    \begin{subfigure}[t]{0.95\textwidth}
    \centering
    \includegraphics[width=\textwidth]{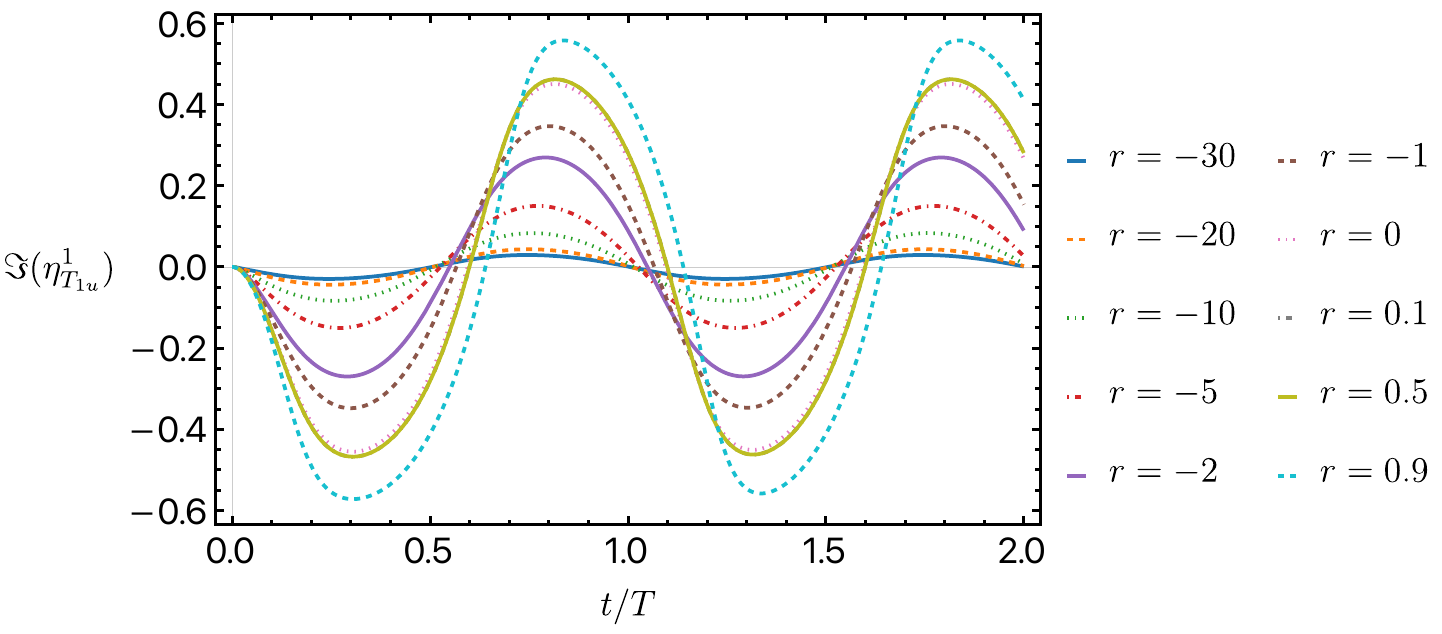}
    \caption{}
    \end{subfigure}
    \caption{Time-dependence (according to~\cref{eq: TDGL A1g,eq: TDGL T1u,eq: TDGL T1u2,eq: TDGL T1u3,eq: TDGL Eg,eq: TDGL Eg2,eq: TDGL T1g,eq: TDGL T1g2,eq: TDGL T1g3,eq: TDGL T2g,eq: TDGL T2g2,eq: TDGL T2g3}) of $\eta_{A_{1g}}$ ($s$-wave) (a) and $\eta_{T_{1u}}^1$ ($p_y$-wave in the 2D limit) (b) OPs for different values of $r_{T_{1u}}$ ranging from $-30$ to $0.9$ under irradiation by linearly polarized uniform microwave.  The TDGL parameters are given in~\cref{tab: Coefficients}, except for $c_{A_{1g}T_{1u}}^2=0$ and values of $r_{T_{1u}}=r$; the dimensionless frequency and amplitude of the microwaves are $\omega=1$ and $A_{0}=0.5$, respectively. We demonstrate that the induced triplet component can be as large as half of the initial value of the $s$-wave component.} 
\label{fig: Plots of OPs vs t for different rT1us}
\end{figure*}

\begin{figure*}[!htbp]
    \centering
    \begin{subfigure}[t]{0.95\textwidth}
    \centering
    \includegraphics[width=\linewidth]{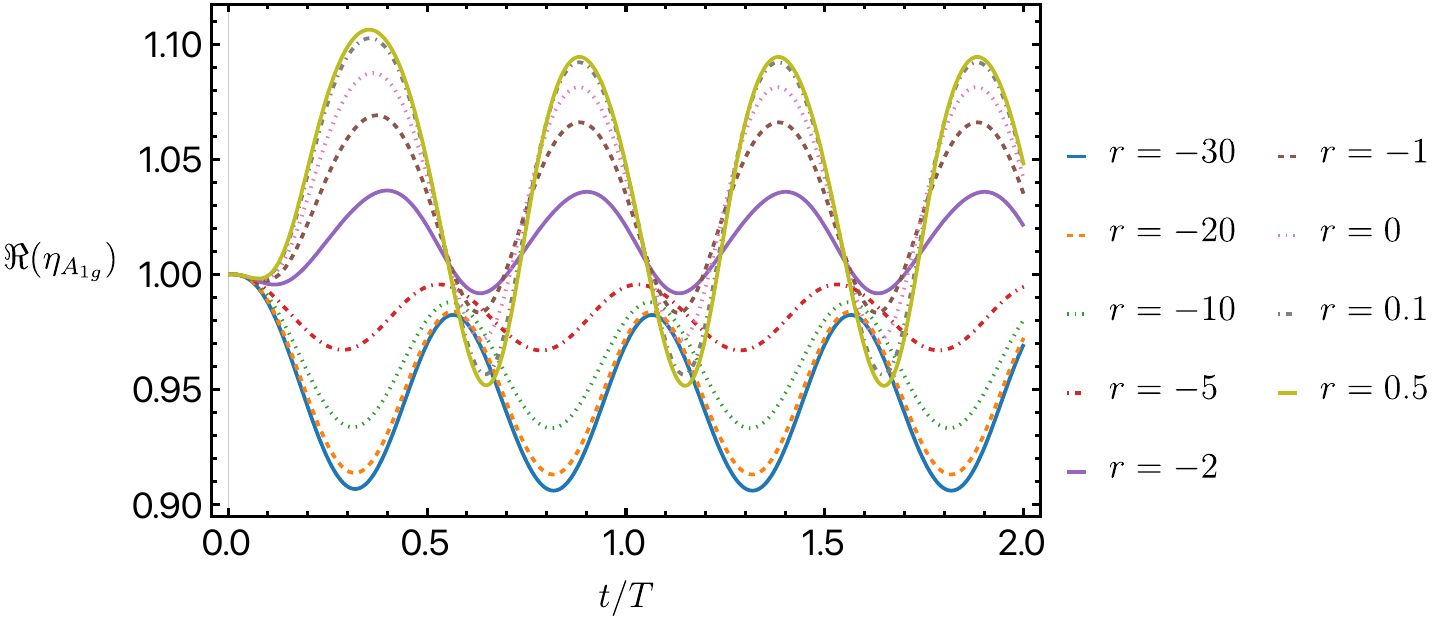}
    \caption{}
    \end{subfigure}
    \begin{subfigure}[t]{0.95\textwidth}
    \centering
    \includegraphics[width=\textwidth]{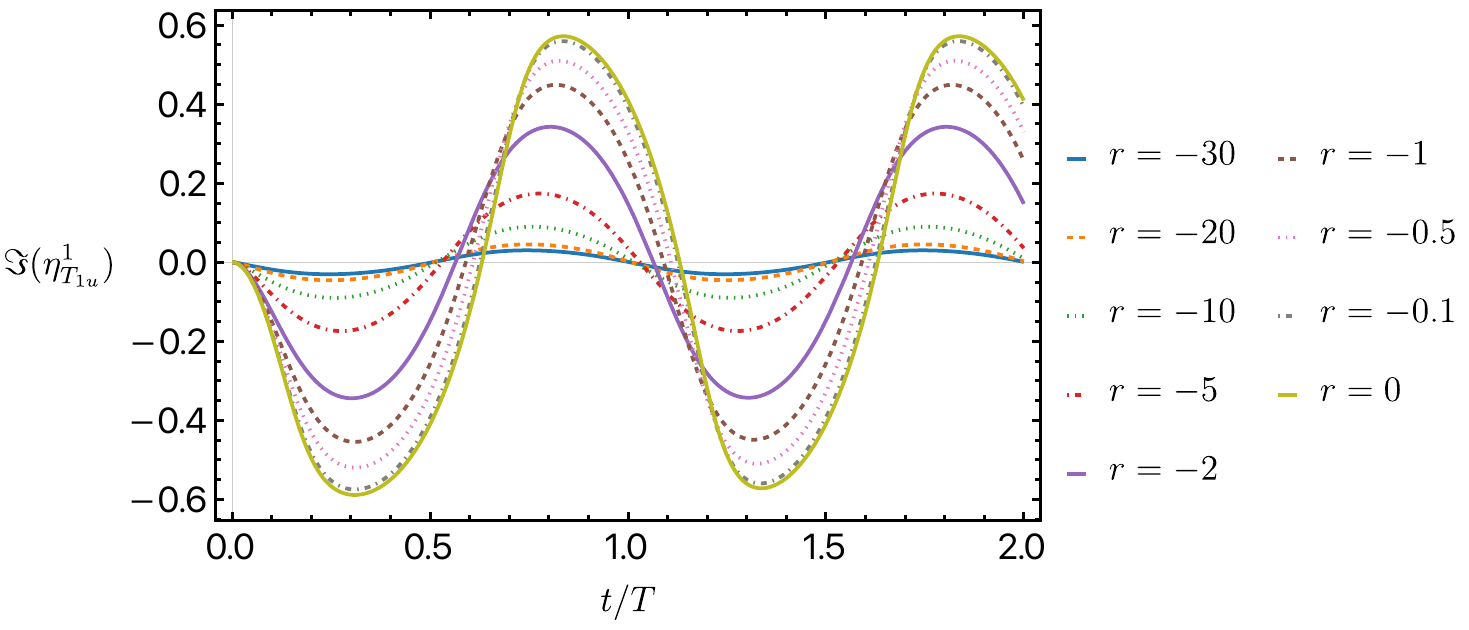}
    \caption{}
    \end{subfigure}
    \caption{Time-dependence (according to~\cref{eq: TDGL A1g,eq: TDGL T1u,eq: TDGL T1u2,eq: TDGL T1u3,eq: TDGL Eg,eq: TDGL Eg2,eq: TDGL T1g,eq: TDGL T1g2,eq: TDGL T1g3,eq: TDGL T2g,eq: TDGL T2g2,eq: TDGL T2g3}) of $\eta_{A_{1g}}$ ($s$-wave) (a) and $\eta_{T_{1u}}^1$ ($p_y$-wave in the 2D limit) (b) OPs for different values of $r_{T_{1u}}$ ranging from $-30$ to $0$ under irradiation by linearly polarized uniform microwave. The TDGL parameters are given in~\cref{tab: Coefficients}, except for $c_{A_{1g}T_{1u}}^2=0$, $c_{A_{1g}T_{1u}}^1=0.1$, and values of $r_{T_{1u}}=r$; the dimensionless frequency and amplitude of the microwaves are $\omega=1$ and $A_{0}=0.5$, respectively. In this case, the induced triplet component can be as large as about a half of the initial value of the $s$-wave component, and the $s$-wave can be enhanced transiently above its initial (equilibrium) value.}
\label{fig: Plots of OPs vs t for different rT1us cA1gT1u=0.1}
\end{figure*}

In~\cref{fig: Plots of OPs vs t for different rT1us,fig: Plots of OPs vs t for different rT1us cA1gT1u=0.1}, we plot the time-dependence of the $\Re(\eta_{A_{1g}})$ and $\Im(\eta_{T_{1u}})$ OPs for different values of $r_{T_{1u}}$ ranging from $-30$ to $0.9$ and from $-30$ to $0$, respectively, for the incident uniform linearly polarized microwave with dimensionless $\omega=1$ and $A_{0x}=0.5$. For these plots, we use the same values of TDGL parameters as before given in~\cref{tab: Coefficients}, except that we set $c_{A_{1g}T_{1u}}^2=0$ for simulations in~\cref{fig: Plots of OPs vs t for different rT1us} to ensure that in the absence of perturbations the ground state is purely $s$-wave for $r_{T_{1u}}>0$, and for data in~\cref{fig: Plots of OPs vs t for different rT1us cA1gT1u=0.1}, we additionally set $c_{A_{1g}T_{1u}}^1=0.1$. As can be seen, for values of $\abs{r_{T_{1u}}} \lesssim 1$, a significant amplitude of the $\eta_{T_{1u}^1}$ OP ($p_y$-wave in the 2D limit), up to about $50$ \% of the initial amplitude of the $s$-wave OP, can be achieved. In such a case, as was pointed out in Ref.~\cite{gassner2024}, 
if the microwave pulse is suitably tailored, the system might relax into the metastable triplet state after the pulse. However, the description of such a scenario within the limits of validity of the TDGL model requires very close competition between the dominant $s$-wave and the subdominant $p$-wave states. As can be seen in~\cref{fig: Plots of OPs vs t for different rT1us cA1gT1u=0.1}, the instantaneous and average values of the $\eta_{A_{1g}}$ OP ($s$-wave) can be increased to values above its equilibrium value. This can be roughly explained as the sum of terms in the r.h.s. of the equation of motion for $\eta_{A_{1g}}$,~\cref{eq: TDGL A1g},
\begin{align*}
2 a_{A_{1g}T_{1u}} i \eta_{T_{1u}}^1 +\left( c_{A_{1g}T_{1u}}^1 \eta_{A_{1g}} - c_{A_{1g}T_{1u}}^2 \eta_{A_{1g}}^* \right) \sum_{i=1}^3 \abs{\eta_{T_{1u}}^i}^2,
\end{align*}
effectively counteract the term $-A_x^2 \eta_{A_{1g}}$, which decreases the effective temperature.

\section{Conclusions and Discussion}
\label{sec:conclusions}

We have shown within the generalized GL framework that triplet OP and lower-symmetry singlet OPs can be generated in an $s$-wave SCs via microwave irradiation. The generation of the triplet component is possible only in the presence of SOC, which enables a linear in spatial derivatives and triplet OP Lifshitz-like invariant. The coefficient in front of this invariant is expected to be approximately linear in the SOC coupling when the SOC is relatively weak~\cite{gassner2024}. After the minimal substitution procedure, this term brings a linear in the vector potential and $s$-wave OP term into the equation of motion for the triplet OP, see~\cref{eq: TDGL T1u,eq: TDGL T1u2,eq: TDGL T1u3}. 

For linearly polarized light, the induced triplet OP oscillates around zero average, whereas the induced singlet OPs acquire a non-zero average rectified component. The non-zero average value of the induced triplet OP can be achieved via irradiation by the circularly polarized beam; see~\cref{fig: Plots of OPs vs t}(d) and~\cref{fig: Plots of OPs vs t circular1,fig: Plots of OPs vs t circular2,fig: Plots of OPs vs t circular3,fig: Plots of OPs vs t circular4}. In this case, the values of the components of the triplet OP oscillate around zeros but do not turn to zero simultaneously.

Presented results further illustrate the utility of the stirring superconducting state with light, the so called quantum printing (QP) approach\cite{aeppli2025quantumprinting}, in which the quantum state or geometrical structure of light modifies the quantum state of matter. Recently, in the context of superconducting condensates, the QP approach was exemplified by the dynamical induction of vortex-antivortex pairs~\cite{yeh2025a,yeh2025b} in $s$-wave superconductors, excitations of Higgs modes in superconductors~\cite{kang2025a,kang2025b}, and the engineering of a dynamical anomalous Josephson junction~\cite{yerzhakov2024} in superconducting circuits via structured light.

Practical implementations of the proposed ideas would require a judicious choice of the superconducting material. Most likely, the $d$-wave high-$T_c$ superconductors could be a good candidate. Given the strict range of microscopic validity of the time-dependent GL theory, 
the current work presents rather a proof of principle that induction of the triplet component in a centrosymmetric singlet SC via microwave radiation is possible. Microscopic calculations are needed to confirm that our findings remain valid at lower temperatures. Potentially, the induction of the triplet OP may open a route to engineering non-equilibrium topologically non-trivial SC states or new ways to control spin-polarized supercurrents.

\begin{acknowledgments}
We gratefully acknowledge helpful discussions with G. Cardoso,  S. Gassner,  S. Karabatsos, I. Khaymovich , A. Talkachov, and T.T. Yeh.
This work was supported by the U.S. Department of Energy, Office of
Science, Office of Basic Energy Sciences under Award No. DE-SC-0025580 (A.V.B., H.Y. concept, calculations, and writing). H.Y. was also supported by the European Research Council under the European Union Seventh Framework ERC-2018-SYG 810451 HERO, the Knut and Alice Wallenberg Foundation KAW 2019.0068 (travel, writing). The development of the code is supported by DOE award No. DE-SC-0025580.
\end{acknowledgments}

\section*{Data Availability}

Data sets are available from the corresponding author on
request. Movies of the simulated dynamics of SC condensate on a 2D film irradiated by linearly and circularly polarized Gaussian beams  and the Python code are available openly in the \textit{GeneralizedTDGL\_Oh} Github repository at 
\url{https://github.com/hennadii-phys/GeneralizedTDGL\_Oh}.

\appendix

\section{Direct product table}
\label{app:producttable}
In this Appendix, we show a direct product table for the $O_h$ point group.

\begin{table*}[!htbp]
\centering
\caption{Direct product table for the $O_h$ point group (for binary products only).}
\renewcommand{\arraystretch}{1.25}
{\tiny
\begin{tabular}{@{}>{$}l<{$}*{10}{>{$}c<{$}}@{}}
\toprule
& A_{1g} & A_{2g} & E_g & T_{1g} & T_{2g} & A_{1u} & A_{2u} & E_u & T_{1u} & T_{2u} \\
\midrule
A_{1g} & A_{1g} & A_{2g} & E_g & T_{1g} & T_{2g} & A_{1u} & A_{2u} & E_u & T_{1u} & T_{2u} \\
A_{2g} & A_{2g} & A_{1g} & E_g & T_{2g} & T_{1g} & A_{2u} & A_{1u} & E_u & T_{2u} & T_{1u} \\
E_g    & E_g    & E_g    & A_{1g}\oplus [A_{2g}]\oplus E_g &
          T_{1g}\oplus T_{2g} &
          T_{1g}\oplus T_{2g} &
          E_u & E_u & A_{1u}\oplus A_{2u}\oplus E_u &
          T_{1u}\oplus T_{2u} &
          T_{1u}\oplus T_{2u} \\
T_{1g} & T_{1g} & T_{2g} & T_{1g}\oplus T_{2g} &
          A_{1g}\oplus E_g\oplus [T_{1g}]\oplus T_{2g} &
          A_{2g}\oplus E_g\oplus T_{1g}\oplus T_{2g} &
          T_{1u} & T_{2u} & T_{1u}\oplus T_{2u} &
          A_{1u}\oplus E_u\oplus T_{1u}\oplus T_{2u} &
          A_{2u}\oplus E_u\oplus T_{1u}\oplus T_{2u} \\
T_{2g} & T_{2g} & T_{1g} & T_{1g}\oplus T_{2g} &
          A_{2g}\oplus E_g\oplus T_{1g}\oplus T_{2g} &
          A_{1g}\oplus E_g\oplus [T_{1g}]\oplus T_{2g} &
          T_{2u} & T_{1u} & T_{1u}\oplus T_{2u} &
          A_{2u}\oplus E_u\oplus T_{1u}\oplus T_{2u} &
          A_{1u}\oplus E_u\oplus T_{1u}\oplus T_{2u} \\
A_{1u} & A_{1u} & A_{2u} & E_u & T_{1u} & T_{2u} &
          A_{1g} & A_{2g} & E_g & T_{1g} & T_{2g} \\
A_{2u} & A_{2u} & A_{1u} & E_u & T_{2u} & T_{1u} &
          A_{2g} & A_{1g} & E_g & T_{2g} & T_{1g} \\
E_u    & E_u    & E_u    & A_{1u}\oplus A_{2u}\oplus E_u &
          T_{1u}\oplus T_{2u} &
          T_{1u}\oplus T_{2u} &
          E_g & E_g & A_{1g}\oplus [A_{2g}]\oplus E_g &
          T_{1g}\oplus T_{2g} &
          T_{1g}\oplus T_{2g} \\
T_{1u} & T_{1u} & T_{2u} & T_{1u}\oplus T_{2u} &
          A_{1u}\oplus E_u\oplus T_{1u}\oplus T_{2u} &
          A_{2u}\oplus E_u\oplus T_{1u}\oplus T_{2u} &
          T_{1g} & T_{2g} & T_{1g}\oplus T_{2g} &
          A_{1g}\oplus E_g\oplus [T_{1g}]\oplus T_{2g} &
          A_{2g}\oplus E_g\oplus T_{1g}\oplus T_{2g} \\
T_{2u} & T_{2u} & T_{1u} & T_{1u}\oplus T_{2u} &
          A_{2u}\oplus E_u\oplus T_{1u}\oplus T_{2u} &
          A_{1u}\oplus E_u\oplus T_{1u}\oplus T_{2u} &
          T_{2g} & T_{1g} & T_{1g}\oplus T_{2g} &
          A_{2g}\oplus E_g\oplus T_{1g}\oplus T_{2g} &
          A_{1g}\oplus E_g\oplus [T_{1g}]\oplus T_{2g} \\
\bottomrule
\end{tabular}
}
\label{tab: Product table}
\end{table*}

\section{Terms in the Lagrangian not shown in the main part}
\label{app:restofthelagrangian}
In this Appendix, we provide the rest of constituent parts of the full static Lagrangian density,~\cref{eq: full static Lagrangian}, not displayed in the main text.

\begin{widetext}
\begin{align}
\mathcal{L}_{T_{1u},E_g}^\nabla
&= a_{T_{1u}E_g}\Big\{
\tfrac12\,(D_x\eta_{T_{1u}}^{\,1})^{*}\,\eta_{E_g}^{\,1}
-\tfrac12\,(D_x\eta_{T_{1u}}^{\,1})^{*}\,\eta_{E_g}^{\,2}
-\tfrac12\,(D_y\eta_{T_{1u}}^{\,2})^{*}\,\eta_{E_g}^{\,1}
-\tfrac12\,(D_y\eta_{T_{1u}}^{\,2})^{*}\,\eta_{E_g}^{\,2}
+(D_z\eta_{T_{1u}}^{\,3})^{*}\,\eta_{E_g}^{\,2}
+\text{c.c.}\Big\}
\\ \nn
&\quad
- \, a_{T_{1u}E_g}\Big\{
\tfrac12\,\eta_{T_{1u}}^{\,1*}\,D_x\eta_{E_g}^{\,1}
-\tfrac12\,\eta_{T_{1u}}^{\,1*}\,D_x\eta_{E_g}^{\,2}
-\tfrac12\,\eta_{T_{1u}}^{\,2*}\,D_y\eta_{E_g}^{\,1}
-\tfrac12\,\eta_{T_{1u}}^{\,2*}\,D_y\eta_{E_g}^{\,2}
+\eta_{T_{1u}}^{\,3*}\,D_z\eta_{E_g}^{\,2}
+\text{c.c.}\Big\}, \\
\mathcal{L}_{T_{1u},T_{1g}}^\nabla
&= a_{T_{1u}T_{1g}}\Big\{
-\eta^{\,1*}_{T_{1u}}\,D_y\eta^{\,1}_{T_{1g}}
-\eta^{\,1*}_{T_{1u}}\,D_z\eta^{\,2}_{T_{1g}}
+\eta^{\,2*}_{T_{1u}}\,D_x\eta^{\,1}_{T_{1g}}
-\eta^{\,2*}_{T_{1u}}\,D_z\eta^{\,3}_{T_{1g}}
+\eta^{\,3*}_{T_{1u}}\,D_x\eta^{\,2}_{T_{1g}}
+\eta^{\,3*}_{T_{1u}}\,D_y\eta^{\,3}_{T_{1g}}
+\text{c.c.}\Big\} \\ \nn
&\quad
- a_{T_{1u}T_{1g}}\Big\{
-(D_y\eta^{\,1}_{T_{1u}})^*\,\eta^{\,1}_{T_{1g}}
-(D_z\eta^{\,1}_{T_{1u}})^*\,\eta^{\,2}_{T_{1g}}
+(D_x\eta^{\,2}_{T_{1u}})^*\,\eta^{\,1}_{T_{1g}} \\ \nn
&\hspace{3.2cm}
-(D_z\eta^{\,2}_{T_{1u}})^*\,\eta^{\,3}_{T_{1g}}
+(D_x\eta^{\,3}_{T_{1u}})^*\,\eta^{\,2}_{T_{1g}}
+(D_y\eta^{\,3}_{T_{1u}})^*\,\eta^{\,3}_{T_{1g}}
+\text{c.c.}\Big\}, \\
\mathcal{L}_{T_{1u},T_{2g}}^\nabla
&= a_{T_{1u}T_{2g}}\Big\{
\eta_{T_{1u}}^{\,1*}\,D_y\eta_{T_{2g}}^{\,1}
+\eta_{T_{1u}}^{\,1*}\,D_z\eta_{T_{2g}}^{\,2}
+\eta_{T_{1u}}^{\,2*}\,D_x\eta_{T_{2g}}^{\,1}
+\eta_{T_{1u}}^{\,2*}\,D_z\eta_{T_{2g}}^{\,3} 
+\eta_{T_{1u}}^{\,3*}\,D_x\eta_{T_{2g}}^{\,2}
+\eta_{T_{1u}}^{\,3*}\,D_y\eta_{T_{2g}}^{\,3}
+\text{c.c.}\Big\} \\ \nn
&\quad
- a_{T_{1u}T_{2g}}\Big\{
(D_y\eta_{T_{1u}}^{\,1})^*\,\eta_{T_{2g}}^{\,1}
+(D_z\eta_{T_{1u}}^{\,1})^*\,\eta_{T_{2g}}^{\,2}
+(D_x\eta_{T_{1u}}^{\,2})^*\,\eta_{T_{2g}}^{\,1} \\ \nn
&\hspace{3.0cm}
+(D_z\eta_{T_{1u}}^{\,2})^*\,\eta_{T_{2g}}^{\,3}
+(D_x\eta_{T_{1u}}^{\,3})^*\,\eta_{T_{2g}}^{\,2}
+(D_y\eta_{T_{1u}}^{\,3})^*\,\eta_{T_{2g}}^{\,3}
+\text{c.c.}\Big\}.
\end{align}
\end{widetext}

\begin{widetext}
\begin{align}
\mathcal{L}_{E_g,T_{1g}}^\nabla
&= a^{1}_{E_gT_{1g}}
\Big\{
-\tfrac13\,D_x\eta^{\,1}_{E_g}\,(D_y\eta^{\,1}_{T_{1g}})^{*}
-\tfrac13\,D_y\eta^{\,1}_{E_g}\,(D_x\eta^{\,1}_{T_{1g}})^{*}
-\tfrac23\,D_x\eta^{\,1}_{E_g}\,(D_z\eta^{\,2}_{T_{1g}})^{*}
+\tfrac13\,D_z\eta^{\,1}_{E_g}\,(D_x\eta^{\,2}_{T_{1g}})^{*} \\ \nn
&\qquad
+\tfrac23\,D_y\eta^{\,1}_{E_g}\,(D_z\eta^{\,3}_{T_{1g}})^{*}
-\tfrac13\,D_z\eta^{\,1}_{E_g}\,(D_y\eta^{\,3}_{T_{1g}})^{*}
+ D_x\eta^{\,2}_{E_g}\,(D_y\eta^{\,1}_{T_{1g}})^{*}
- D_y\eta^{\,2}_{E_g}\,(D_x\eta^{\,1}_{T_{1g}})^{*} \\ \nn
&\qquad
+ D_z\eta^{\,2}_{E_g}\,(D_x\eta^{\,2}_{T_{1g}})^{*}
+ D_z\eta^{\,2}_{E_g}\,(D_y\eta^{\,3}_{T_{1g}})^{*}
+\text{c.c.}
\Big\}
\\[4pt] \nn
&\quad
+ a^{2}_{E_gT_{1g}}
\Big\{
-\tfrac13\,D_x\eta^{\,1}_{E_g}\,(D_y\eta^{\,1}_{T_{1g}})^*
-\tfrac13\,D_y\eta^{\,1}_{E_g}\,(D_x\eta^{\,1}_{T_{1g}})^*
+\tfrac13\,D_x\eta^{\,1}_{E_g}\,(D_z\eta^{\,2}_{T_{1g}})^*
-\tfrac23\,D_z\eta^{\,1}_{E_g}\,(D_x\eta^{\,2}_{T_{1g}})^* \\ \nn
&\qquad
-\tfrac13\,D_y\eta^{\,1}_{E_g}\,(D_z\eta^{\,3}_{T_{1g}})^*
+\tfrac23\,D_z\eta^{\,1}_{E_g}\,(D_y\eta^{\,3}_{T_{1g}})^*
- D_x\eta^{\,2}_{E_g}\,(D_y\eta^{\,1}_{T_{1g}}))^*
+ D_y\eta^{\,2}_{E_g}\,(D_x\eta^{\,1}_{T_{1g}})^* \\ \nn
&\qquad
+ D_x\eta^{\,2}_{E_g}\,(D_z\eta^{\,2}_{T_{1g}})^*
+ D_y\eta^{\,2}_{E_g}\,(D_z\eta^{\,3}_{T_{1g}})^*
+\text{c.c.}
\Big\}, \\
\mathcal{L}_{E_g,T_{2g}}^\nabla
&= a^{1}_{E_gT_{2g}}
\Big\{
\tfrac13\,D_x\eta_{E_g}^{\,1}\,(D_y\eta_{T_{2g}}^{\,1})^{*}
-\tfrac13\,D_y\eta_{E_g}^{\,1}\,(D_x\eta_{T_{2g}}^{\,1})^{*}
+\tfrac23\,D_x\eta_{E_g}^{\,1}\,(D_z\eta_{T_{2g}}^{\,2})^{*}
+\tfrac13\,D_z\eta_{E_g}^{\,1}\,(D_x\eta_{T_{2g}}^{\,2})^{*} \\ \nn
&\qquad
-\tfrac23\,D_y\eta_{E_g}^{\,1}\,(D_z\eta_{T_{2g}}^{\,3})^{*}
-\tfrac13\,D_z\eta_{E_g}^{\,1}\,(D_y\eta_{T_{2g}}^{\,3})^{*}
- D_x\eta_{E_g}^{\,2}\,(D_y\eta_{T_{2g}}^{\,1})^{*}
- D_y\eta_{E_g}^{\,2}\,(D_x\eta_{T_{2g}}^{\,1})^{*} \\ \nn
&\qquad
+ D_z\eta_{E_g}^{\,2}\,(D_x\eta_{T_{2g}}^{\,2})^{*}
+ D_z\eta_{E_g}^{\,2}\,(D_y\eta_{T_{2g}}^{\,3})^{*}
+ \text{c.c.}
\Big\}
\\[2pt] \nn
&\quad
+ a^{2}_{E_gT_{2g}}
\Big\{
-\tfrac13\,D_x\eta_{E_g}^{\,1}\,(D_y\eta_{T_{2g}}^{\,1})^*
+\tfrac13\,D_y\eta_{E_g}^{\,1}\,(D_x\eta_{T_{2g}}^{\,1})^*
+\tfrac13\,D_x\eta_{E_g}^{\,1}\,(D_z\eta_{T_{2g}}^{\,2})^*
+\tfrac23\,D_z\eta_{E_g}^{\,1}\,(D_x\eta_{T_{2g}}^{\,2})^* \\ \nn
&\qquad
-\tfrac13\,D_y\eta_{E_g}^{\,1}\,(D_z\eta_{T_{2g}}^{\,3})^*
-\tfrac23\,D_z\eta_{E_g}^{\,1}\,(D_y\eta_{T_{2g}}^{\,3})^*
- D_x\eta_{E_g}^{\,2}\,(D_y\eta_{T_{2g}}^{\,1})^*
- D_y\eta_{E_g}^{\,2}\,(D_x\eta_{T_{2g}}^{\,1})^* \\ \nn
&\qquad
+ D_x\eta_{E_g}^{\,2}\,(D_z\eta_{T_{2g}}^{\,2})^*
+ D_y\eta_{E_g}^{\,2}\,(D_z\eta_{T_{2g}}^{\,3})^*
+ \text{c.c.}
\Big\},
\end{align}

\begin{align}
\mathcal{L}_{T_{1g},T_{2g}}^\nabla
&= a^{1}_{T_{1g}T_{2g}}
\Big\{
-\,D_x\eta^{\,1}_{T_{1g}}\,(D_x\eta^{\,1}_{T_{2g}})^{*}
+\,D_y\eta^{\,1}_{T_{1g}}\,(D_y\eta^{\,1}_{T_{2g}})^{*}
-\,D_x\eta^{\,2}_{T_{1g}}\,(D_x\eta^{\,2}_{T_{2g}})^{*}
+\,D_z\eta^{\,2}_{T_{1g}}\,(D_z\eta^{\,2}_{T_{2g}})^{*} \\ \nn
&\qquad
-\,D_y\eta^{\,3}_{T_{1g}}\,(D_y\eta^{\,3}_{T_{2g}})^{*}
+\,D_z\eta^{\,3}_{T_{1g}}\,(D_z\eta^{\,3}_{T_{2g}})^{*}
+\text{c.c.}
\Big\} \\[4pt] \nn
&\quad
+ a^{2}_{T_{1g}T_{2g}}
\Big\{
-\,D_x\eta^{\,1}_{T_{1g}}\,(D_z\eta^{\,3}_{T_{2g}})^{*}
+\,D_z\eta^{\,3}_{T_{1g}}\,(D_x\eta^{\,1}_{T_{2g}})^{*}
+\,D_y\eta^{\,1}_{T_{1g}}\,(D_z\eta^{\,2}_{T_{2g}})^{*}
+\,D_z\eta^{\,2}_{T_{1g}}\,(D_y\eta^{\,1}_{T_{2g}})^{*} \\ \nn
&\qquad
-\,D_x\eta^{\,2}_{T_{1g}}\,(D_y\eta^{\,3}_{T_{2g}})^{*}
+\,D_z\eta^{\,3}_{T_{1g}}\,(D_x\eta^{\,1}_{T_{2g}})^{*}
+\text{c.c.}
\Big\} \\[4pt] \nn
&\quad
+ a^{3}_{T_{1g}T_{2g}}
\Big\{
-\,D_z\eta^{\,1}_{T_{1g}}\,(D_x\eta^{\,3}_{T_{2g}})^{*}
+\,D_x\eta^{\,3}_{T_{1g}}\,(D_z\eta^{\,1}_{T_{2g}})^{*}
+\,D_z\eta^{\,1}_{T_{1g}}\,(D_y\eta^{\,2}_{T_{2g}})^{*}
+\,D_y\eta^{\,2}_{T_{1g}}\,(D_z\eta^{\,1}_{T_{2g}})^{*} \\ \nn
&\qquad
-\,D_y\eta^{\,2}_{T_{1g}}\,(D_x\eta^{\,3}_{T_{2g}})^{*}
-\,D_x\eta^{\,3}_{T_{1g}}\,(D_y\eta^{\,2}_{T_{2g}})^{*}
+\text{c.c.}
\Big\}.
\end{align}
\end{widetext}

\section{TDGL equations not shown in the main text}
\label{app:restofthetdlgeqs}
 In this Appendix, we write down the rest of the TDGL equations not shown in the main text:
\begin{widetext}
\begin{align}
\label{eq: TDGL T1g}
\Gamma_{T_{1g}}\,D_t \eta^{\,1}_{T_{1g}}
&= r_{T_{1g}}\,\eta^{\,1}_{T_{1g}}
+ a^{1}_{T_{1g}}\,D_x^{2}\eta^{\,1}_{T_{1g}}
+ a^{2}_{T_{1g}}\,D_y^{2}\eta^{\,1}_{T_{1g}}
+ a^{3}_{T_{1g}}\,D_xD_y\,\eta^{\,2}_{T_{1g}}
+ a^{4}_{T_{1g}}\,D_yD_x\,\eta^{\,2}_{T_{1g}}
\\ \nn
&\quad
-2 b^{1}_{T_{1g}}\!\left(\sum_{i=1}^{3}\lvert\eta^{\,i}_{T_{1g}}\rvert^{2}\right)\eta^{\,1}_{T_{1g}}
-2 b^{2}_{T_{1g}}\!\left(\sum_{i=1}^{3}\big(\eta^{\,i}_{T_{1g}}\big)^{2}\right)\eta^{\,1*}_{T_{1g}}
- b^{3}_{T_{1g}}\!\left(\lvert\eta^{\,2}_{T_{1g}}\rvert^{2}
+\lvert\eta^{\,3}_{T_{1g}}\rvert^{2}\right)\eta^{\,1}_{T_{1g}}
\\ \nn
&\quad
+ a_{A_{1g}T_{1g}}\!\left(-D_yD_x\,\eta_{A_{1g}}+D_xD_y\,\eta_{A_{1g}}\right)
+ 2a_{T_{1u}T_{1g}}\!\left(-D_y\eta^{\,1}_{T_{1u}}+D_x\eta^{\,2}_{T_{1u}}\right)
\\ \nn
&\quad
+ a^{1}_{E_gT_{1g}}\!\left(-\tfrac13 D_yD_x\,\eta^{\,1}_{E_g}
-\tfrac13 D_xD_y\,\eta^{\,1}_{E_g}
+ D_yD_x\,\eta^{\,2}_{E_g}-D_xD_y\,\eta^{\,2}_{E_g}\right)
\\ \nn
&\quad
+ a^{2}_{E_gT_{1g}}\!\left(-\tfrac13 D_yD_x\,\eta^{\,1}_{E_g}
-\tfrac13 D_xD_y\,\eta^{\,1}_{E_g}
- D_yD_x\,\eta^{\,2}_{E_g}+D_xD_y\,\eta^{\,2}_{E_g}\right)
\\ \nn
&\quad
+ a^{1}_{T_{1g}T_{2g}}\!\left(-D_x^{2}\eta^{\,1}_{T_{2g}}
+ D_y^{2}\eta^{\,1}_{T_{2g}}\right)\\ \nn
&\quad -c_{A_{1g}T_{1g}}^1 \abs{\eta_{A_{1g}}}^2 \eta_{T_{1g}}^1 - c_{A_{1g}T_{1g}}^2 \left(\eta_{A_{1g}}\right)^2 \eta_{T_{1g}}^{1*},
\end{align}
\begin{align}
\label{eq: TDGL T1g2}
\Gamma_{T_{1g}}\,D_t \eta^{\,2}_{T_{1g}}
&= r_{T_{1g}}\,\eta^{\,2}_{T_{1g}}
+ a^{1}_{T_{1g}}\,D_y^{2}\eta^{\,2}_{T_{1g}}
+ a^{2}_{T_{1g}}\,D_x^{2}\eta^{\,2}_{T_{1g}}
+ a^{3}_{T_{1g}}\,D_yD_x\,\eta^{\,1}_{T_{1g}}
+ a^{4}_{T_{1g}}\,D_xD_y\,\eta^{\,1}_{T_{1g}}
\\ \nn
&\quad
-2 b^{1}_{T_{1g}}\!\left(\sum_{i=1}^{3}\big|\eta^{\,i}_{T_{1g}}\big|^{2}\right)\eta^{\,2}_{T_{1g}}
-2 b^{2}_{T_{1g}}\!\left(\sum_{i=1}^{3}\big(\eta^{\,i}_{T_{1g}}\big)^{2}\right)\eta^{\,2*}_{T_{1g}}
- b^{3}_{T_{1g}}\!\left(\big|\eta^{\,1}_{T_{1g}}\big|^{2}+\big|\eta^{\,3}_{T_{1g}}\big|^{2}\right)\eta^{\,2}_{T_{1g}}
\\ \nn
&\quad
+ 2 a_{T_{1u}T_{1g}}\,D_x\,\eta^{\,3}_{T_{1u}}
- a^{1}_{T_{1g}T_{2g}}\,D_x^{2}\eta^{\,2}_{T_{2g}}
- a^{2}_{T_{1g}T_{2g}}\,D_xD_y\,\eta^{\,3}_{T_{2g}}
- a^{3}_{T_{1g}T_{2g}}\,D_yD_x\,\eta^{\,3}_{T_{2g}}\\ \nn
&\quad -c_{A_{1g}T_{1g}}^1 \abs{\eta_{A_{1g}}}^2 \eta_{T_{1g}}^2 - c_{A_{1g}T_{1g}}^2 \left(\eta_{A_{1g}}\right)^2 \eta_{T_{1g}}^{2*},
\end{align}
\begin{align}
\label{eq: TDGL T1g3}
\Gamma_{T_{1g}}\,D_t \eta^{\,3}_{T_{1g}}
&= r_{T_{1g}}\,\eta^{\,3}_{T_{1g}}
+ a^{2}_{T_{1g}}\,\left(D_x^{2}\eta^{\,3}_{T_{1g}}
+D_y^{2}\eta^{\,3}_{T_{1g}} \right)
\\ \nn
&\quad
-2 b^{1}_{T_{1g}}\!\left(\sum_{i=1}^{3}\big|\eta^{\,i}_{T_{1g}}\big|^{2}\right)\eta^{\,3}_{T_{1g}}
-2 b^{2}_{T_{1g}}\!\left(\sum_{i=1}^{3}\big(\eta^{\,i}_{T_{1g}}\big)^{2}\right)\eta^{\,3*}_{T_{1g}}
- b^{3}_{T_{1g}}\!\left(\big|\eta^{\,1}_{T_{1g}}\big|^{2}
+\big|\eta^{\,2}_{T_{1g}}\big|^{2}\right)\eta^{\,3}_{T_{1g}}
\\ \nn
&\quad
- a^{1}_{T_{1g}T_{2g}}\,D_y^{2}\eta^{\,3}_{T_{2g}}
- a^{3}_{T_{1g}T_{2g}}\,D_xD_y\,\eta^{\,2}_{T_{2g}}\\ \nn
&\quad -c_{A_{1g}T_{1g}}^1 \abs{\eta_{A_{1g}}}^2 \eta_{T_{1g}}^3 - c_{A_{1g}T_{1g}}^2 \left(\eta_{A_{1g}}\right)^2 \eta_{T_{1g}}^{3*},
\end{align}
for the components of the $g$-wave OP,
\begin{align}
\label{eq: TDGL T2g}
\Gamma_{T_{2g}}\,D_t \eta^{\,1}_{T_{2g}}
&= r_{T_{2g}}\,\eta^{\,1}_{T_{2g}}
+ a^{1}_{T_{2g}}\,D_x^{2}\eta^{\,1}_{T_{2g}}
+ a^{2}_{T_{2g}}\,D_y^{2}\eta^{\,1}_{T_{2g}}
+ a^{3}_{T_{2g}}\,D_xD_y\,\eta^{\,2}_{T_{2g}}
+ a^{4}_{T_{2g}}\,D_yD_x\,\eta^{\,2}_{T_{2g}}
\\ \nn
&\quad
-2 b^{1}_{T_{2g}}\!\left(\sum_{i=1}^{3}\lvert\eta^{\,i}_{T_{2g}}\rvert^{2}\right)\eta^{\,1}_{T_{2g}}
-2 b^{2}_{T_{2g}}\!\left(\sum_{i=1}^{3}\big(\eta^{\,i}_{T_{2g}}\big)^{2}\right)\eta^{\,1*}_{T_{2g}}
- b^{3}_{T_{2g}}\!\left(\lvert\eta^{\,2}_{T_{2g}}\rvert^{2}
+\lvert\eta^{\,3}_{T_{2g}}\rvert^{2}\right)\eta^{\,1}_{T_{2g}}
\\ \nn
&\quad
+ a_{A_{1g}T_{2g}}\!\left(D_yD_x\,\eta_{A_{1g}}+D_xD_y\,\eta_{A_{1g}}\right)
+ 2a_{T_{1u}T_{2g}}\!\left(D_y\eta^{\,1}_{T_{1u}}+D_x\eta^{\,2}_{T_{1u}}\right)
\\ \nn
&\quad
+ a^{1}_{E_gT_{2g}}\!\left(\tfrac13 D_yD_x\,\eta^{\,1}_{E_g}
-\tfrac13 D_xD_y\,\eta^{\,1}_{E_g}
- D_yD_x\,\eta^{\,2}_{E_g}
- D_xD_y\,\eta^{\,2}_{E_g}\right)
\\ \nn
&\quad
+ a^{2}_{E_gT_{2g}}\!\left(-\tfrac13 D_yD_x\,\eta^{\,1}_{E_g}
+\tfrac13 D_xD_y\,\eta^{\,1}_{E_g}
- D_yD_x\,\eta^{\,2}_{E_g}
- D_xD_y\,\eta^{\,2}_{E_g}\right)
\\ \nn
&\quad
+ a^{1}_{T_{1g}T_{2g}}\!\left(-D_x^{2}\eta^{\,1}_{T_{1g}}
+ D_y^{2}\eta^{\,1}_{T_{1g}}\right) -c_{A_{1g}T_{2g}}^1 \abs{\eta_{A_{1g}}}^2 \eta_{T_{2g}}^1 - c_{A_{1g}T_{2g}}^2 \left(\eta_{A_{1g}}\right)^2 \eta_{T_{2g}}^{1*},
\end{align}
\begin{align}
\label{eq: TDGL T2g2}
\Gamma_{T_{2g}}\,D_t \eta^{\,2}_{T_{2g}}
&= r_{T_{2g}}\,\eta^{\,2}_{T_{2g}}
+ a^{1}_{T_{2g}}\,D_y^{2}\eta^{\,2}_{T_{2g}}
+ a^{2}_{T_{2g}}\,D_x^{2}\eta^{\,2}_{T_{2g}}
+ a^{3}_{T_{2g}}\,D_yD_x\,\eta^{\,1}_{T_{2g}}
+ a^{4}_{T_{2g}}\,D_xD_y\,\eta^{\,1}_{T_{2g}}
\\ \nn
&\quad
-2 b^{1}_{T_{2g}}\!\left(\sum_{i=1}^{3}\big|\eta^{\,i}_{T_{2g}}\big|^{2}\right)\eta^{\,2}_{T_{2g}}
-2 b^{2}_{T_{2g}}\!\left(\sum_{i=1}^{3}\big(\eta^{\,i}_{T_{2g}}\big)^{2}\right)\eta^{\,2*}_{T_{2g}}
- b^{3}_{T_{2g}}\!\left(\big|\eta^{\,1}_{T_{2g}}\big|^{2}
+\big|\eta^{\,3}_{T_{2g}}\big|^{2}\right)\eta^{\,2}_{T_{2g}}
\\ \nn
&\quad
+ 2 a_{T_{1u}T_{2g}}\,D_x\,\eta^{\,3}_{T_{1u}}
- a^{1}_{T_{1g}T_{2g}}\,D_x^{2}\eta^{\,2}_{T_{1g}}
- a^{3}_{T_{1g}T_{2g}}\,D_yD_x\,\eta^{\,3}_{T_{1g}}\\ \nn
&\quad -c_{A_{1g}T_{2g}}^1 \abs{\eta_{A_{1g}}}^2 \eta_{T_{2g}}^2 - c_{A_{1g}T_{2g}}^2 \left(\eta_{A_{1g}}\right)^2 \eta_{T_{2g}}^{2*},
\end{align}
\begin{align}
\label{eq: TDGL T2g3}
\Gamma_{T_{2g}}\,D_t \eta^{\,3}_{T_{2g}}
&= r_{T_{2g}}\,\eta^{\,3}_{T_{2g}}
+ a^{2}_{T_{2g}}\!\left(D_x^{2}+D_y^{2}\right)\eta^{\,3}_{T_{2g}}
\\ \nn
&\quad
- 2 b^{1}_{T_{2g}}\!\left(\sum_{i=1}^{3}\big|\eta^{\,i}_{T_{2g}}\big|^{2}\right)\eta^{\,3}_{T_{2g}}
- 2 b^{2}_{T_{2g}}\!\left(\sum_{i=1}^{3}\big(\eta^{\,i}_{T_{2g}}\big)^2 \right)\eta^{\,3*}_{T_{2g}}
- b^{3}_{T_{2g}}\!\left(\big|\eta^{\,1}_{T_{2g}}\big|^{2}+\big|\eta^{\,2}_{T_{2g}}\big|^{2}\right)\eta^{\,3}_{T_{2g}}
\\ \nn
&\quad
- a^{1}_{T_{1g}T_{2g}}\,D_y^{2}\eta^{\,3}_{T_{1g}}
- a^{2}_{T_{1g}T_{2g}}\,D_yD_x\,\eta^{\,2}_{T_{1g}}
- a^{3}_{T_{1g}T_{2g}}\,D_xD_y\,\eta^{\,2}_{T_{1g}}\\ \nn
&\quad -c_{A_{1g}T_{2g}}^1 \abs{\eta_{A_{1g}}}^2 \eta_{T_{2g}}^3 - c_{A_{1g}T_{2g}}^2 \left(\eta_{A_{1g}}\right)^2 \eta_{T_{2g}}^{3*},
\end{align}
for the $T_{2g}$ components ($d_{xy}$, $d_{xz}$, and $d_{yz}$) of the $d$-wave OP.
\end{widetext}

\section{Coefficients for the condensed form of the Lagrangian}
\label{app:coefficients}

In this Appendix, we provide values for the coefficients $Q_{\mu \nu, ij}^{\alpha \beta}$ and $P_{\nu, i}^{\alpha \beta}$ in~\cref{eq: Q parts,eq: P parts}:

\begin{widetext}
\begin{align}
Q_{T_{2g}T_{2g},ii}^{ii}
  &= a_{T_{2g}}^{1},\,Q_{T_{2g}T_{2g},11}^{22}
   = a_{T_{2g}}^{2}=\,Q_{T_{2g}T_{2g},11}^{33} =\,
Q_{T_{2g}T_{2g},22}^{11}=\,Q_{T_{2g}T_{2g},22}^{33},\,
Q_{T_{2g}T_{2g},12}^{12}= a_{T_{2g}}^{3},\,\\ \nn
Q_{T_{2g}T_{2g},12}^{21}&= a_{T_{2g}}^{4},\qquad 
\mathrm{else }\,\,\, Q_{T_{2g}T_{2g},ij}^{\alpha\beta}=0,\\ \nn
Q_{T_{1g}T_{2g},11}^{11}
  &= -a_{T_{1g}T_{2g}}^{1}=\,-Q_{T_{1}T_{2g},22}^{11}
   =\,Q_{T_{1g}T_{2g},11}^{22} =\,
Q_{T_{1g}T_{2g},22}^{33},\,Q_{T_{1g}T_{2g},12}^{23}= -a_{T_{1g}T_{2g}}^{2},\,\\ \nn
Q_{T_{1g}T_{2g},21}^{23}&= -a_{T_{1g}T_{2g}}^{3}=\,Q_{T_{1g}T_{2g},12}^{32},\qquad 
\mathrm{else }\,\,\, Q_{T_{1g}T_{2g},ij}^{\alpha\beta}=0,\\ \nn
Q_{T_{1g}T_{1g},ii}^{ii}
  &= a_{T_{1g}}^{1},\,Q_{T_{1g}T_{1g},11}^{22}
   = a_{T_{1g}}^{2}=\,Q_{T_{1g}T_{1g},11}^{33} =\,
Q_{T_{1g}T_{1g},22}^{11}=\,Q_{T_{1g}T_{1g},22}^{33},\,
Q_{T_{1g}T_{1g},12}^{12}= a_{T_{1g}}^{3},\,\\ \nn
Q_{T_{1g}T_{1g},12}^{21}&= a_{T_{1g}}^{4},\qquad 
\mathrm{else }\,\,\, Q_{T_{1g}T_{1g},ij}^{\alpha\beta}=0,\\[6pt] \nn
Q_{E_g T_{2g},12}^{11}
  &= \frac{1}{3}\!\left(a_{E_gT_{2g}}^{1}-a_{E_gT_{2g}}^{2}\right)=\,-Q_{E_g T_{2g},21}^{11},\,
Q_{E_g T_{2g},12}^{21}= -\left(a_{E_gT_{2g}}^{1}+a_{E_gT_{2g}}^{2}\right)
=\, Q_{E_g T_{2g},21}^{21},\qquad \mathrm{else }\,\,\, Q_{E_gT_{2g},ij}^{\alpha\beta}=0,
\\ \nn
Q_{E_g T_{1g},12}^{11}
  &= -\frac{1}{3}\!\left(a_{E_gT_{1g}}^{1}-a_{E_gT_{1g}}^{2}\right)=\,Q_{E_g T_{1g},21}^{11},\,
Q_{E_g T_{1g},12}^{21}=a_{E_gT_{1g}}^{1}-a_{E_gT_{1g}}^{2}
=\, -Q_{E_g T_{1g},21}^{21},\qquad \mathrm{else }\,\,\, Q_{E_gT_{1g},ij}^{\alpha\beta}=0,\\[6pt] \nn
Q_{E_gE_g,11}^{11} &= a_{E_g}^{1}-a_{E_g}^{2} =\,Q_{E_gE_g,22}^{11},\,
Q_{E_gE_g,11}^{22} = a_{E_g}^{1}+a_{E_g}^{2}=\, Q_{E_gE_g,22}^{22},\,
Q_{E_gE_g,11}^{12} = \sqrt{3}\,a_{E_g}^{2}
  =\, Q_{E_gE_g,11}^{21},\\ \nn
Q_{E_gE_g,22}^{12} &= -\sqrt{3}\,a_{E_g}^{2}
  = Q_{E_gE_g,22}^{21},\qquad \mathrm{else }\,\,\, Q_{E_gE_g,ij}^{\alpha\beta}=0,
\\[6pt] \nn
Q_{A_{1g}A_{1g},ij}^{11} &= a_{A_{1g}} \delta_{ij} \delta_{\alpha \beta},\\ \nn
Q_{A_{1g}E_g,11}^{11}
  &= \frac{1}{2}\,a_{A_{1g}E_g}=\,-Q_{A_{1g}E_g,11}^{12}
  =\, -Q_{A_{1g}E_g,22}^{11}=\, - Q_{A_{1g}E_g,22}^{12},\qquad \mathrm{else }\,\,\, Q_{A_{1g}E_g,ij}^{\alpha\beta}=0,
\\ \nn
Q_{A_{1g}T_{2g},12}^{11}
  &= \,a_{A_{1g}T_{2g}} =\,Q_{A_{1g}T_{2g},21}^{11},\qquad \mathrm{else }\,\,\, Q_{A_{1g}T_{2g},ij}^{\alpha\beta}=0,\\ \nn
Q_{A_{1g}T_{1g},12}^{11}
  &= \,-a_{A_{1g}T_{1g}} =\,-Q_{A_{1g}T_{1g},21}^{11},\qquad \mathrm{else }\,\,\, Q_{A_{1g}T_{1g},ij}^{\alpha\beta}=0.
\end{align}

\begin{align}
P_{A_{1g},i}^{\,i1} &= a_{A_{1g}T_{1u}},
\qquad \text{else } P_{A_{1g},i}^{\,\alpha \beta}=0,
\\ \nn
P_{E_g,1}^{\,11} &= \tfrac{1}{2}\,a_{T_{1u}E_g}=\,-P_{E_g,1}^{\,12}=-\,P_{E_g,2}^{\,21}=\,-P_{E_g,2}^{\,22},
\qquad \text{else } P_{E_g,i}^{\,\alpha \beta}=0,
\\ \nn
P_{T_{2g},2}^{\,11} &=\, -a_{T_{1u}T_{2g}}=\,P_{T_{2g},1}^{\,21}=\,P_{T_{2g},1}^{\,32}=\, P_{T_{2g},2}^{\,33},
\qquad \text{else } P_{T_{2g},i}^{\,\alpha \beta}=0,
\\ \nn
P_{T_{1g},2}^{\,11} &=\,a_{T_{1u}T_{2g}}=\,-P_{T_{1g},1}^{\,21}=\,-P_{T_{1g},1}^{\,32}=\, -P_{T_{1g},2}^{\,33},
\qquad \text{else } P_{T_{1g},i}^{\,\alpha \beta}=0.
\end{align}

\end{widetext}

\section{Density plots for the time evolution of OPs}
\label{app:densityplots}

\begin{figure*}[!htbp]
    \includegraphics[width=\linewidth]{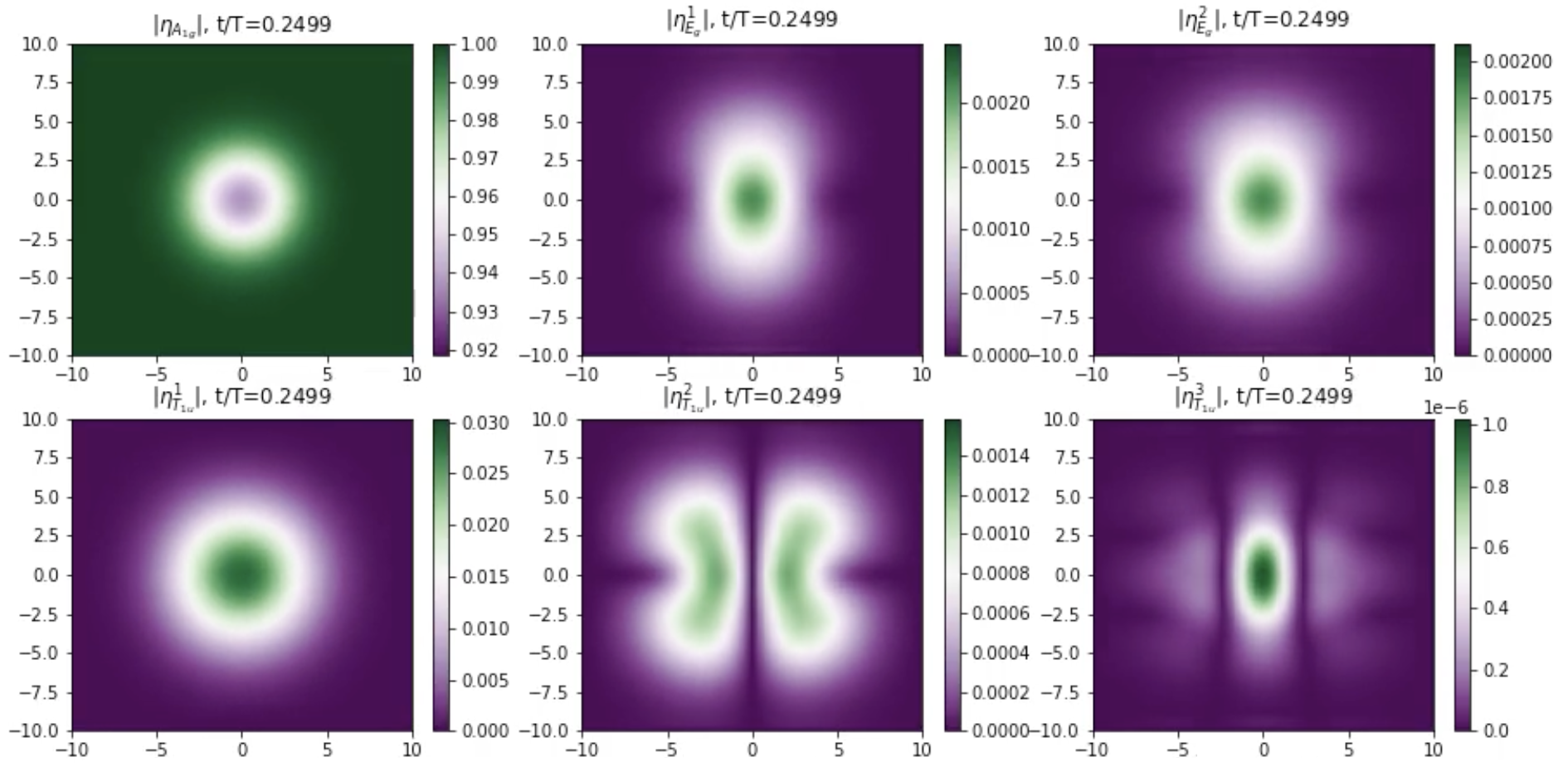}
    \caption{Density plots for the absolute values of $s$-wave $\eta_{A_{1g}}$, $d$-wave $\eta_{Eg}^{1,2}$, and $p$-wave $\eta_{T_{1u}}^{1,2,3}$ OPs under incident linearly polarized microwave beam at approximately quarter of the microwave cycle, $t/T\approx 0.25$, where $T$ is the period of microwave. Coordinates on the thin film are given in units of $\xi$.} 
\label{fig: Plots of OPs vs t linear1}
\end{figure*}

\begin{figure*}[!htbp]
    \includegraphics[width=\linewidth]{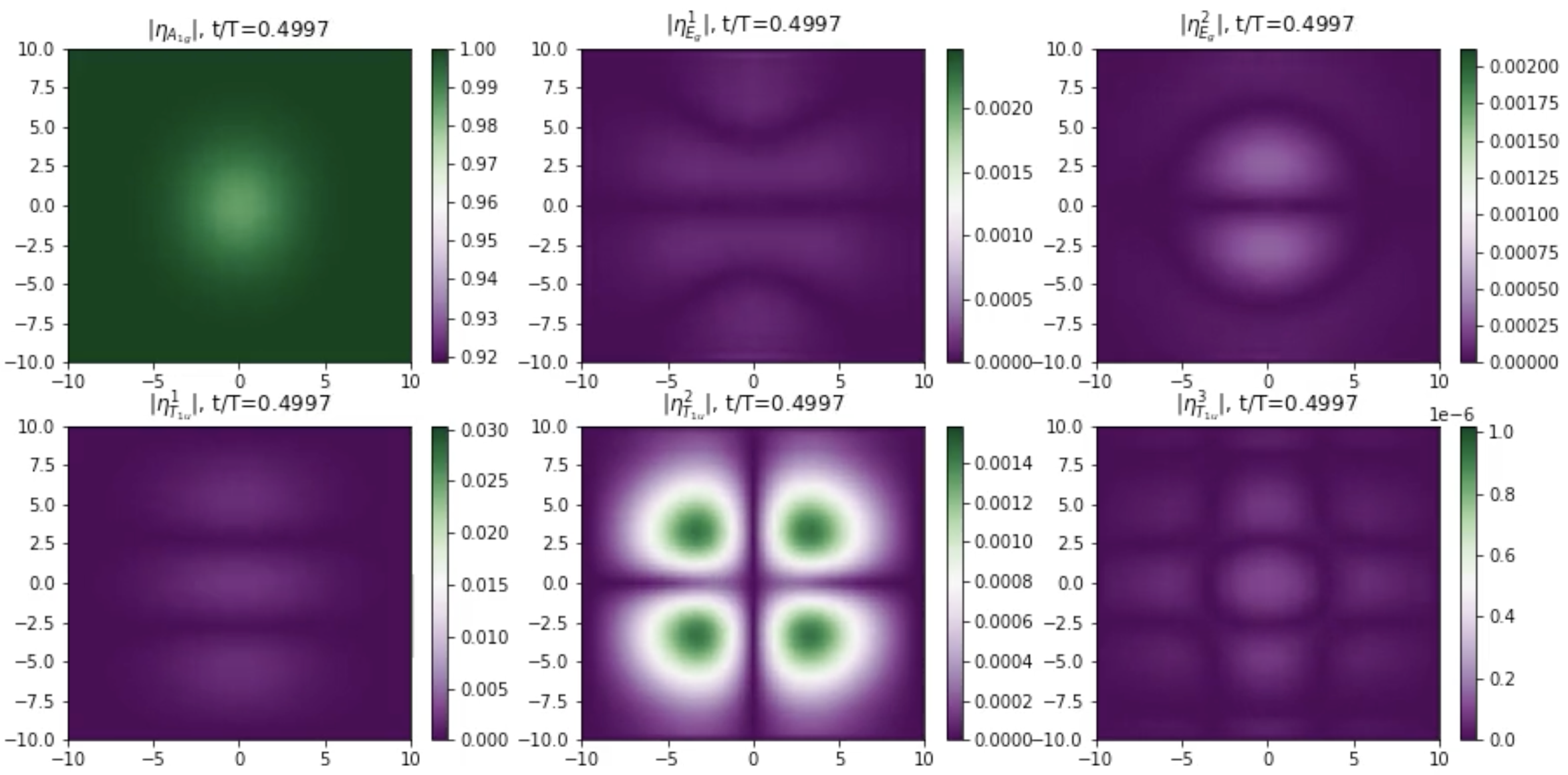}
    \caption{Density plots for the absolute values of $s$-wave $\eta_{A_{1g}}$, $d$-wave $\eta_{Eg}^{1,2}$, and $p$-wave $\eta_{T_{1u}}^{1,2,3}$ OPs under incident linearly polarized microwave beam at approximately half of the microwave cycle, $t/T\approx 0.5$, where $T$ is the period of microwave. Coordinates on the thin film are given in units of $\xi$.} 
\label{fig: Plots of OPs vs t linear2}
\end{figure*}

\begin{figure*}[!htbp]
    \includegraphics[width=\linewidth]{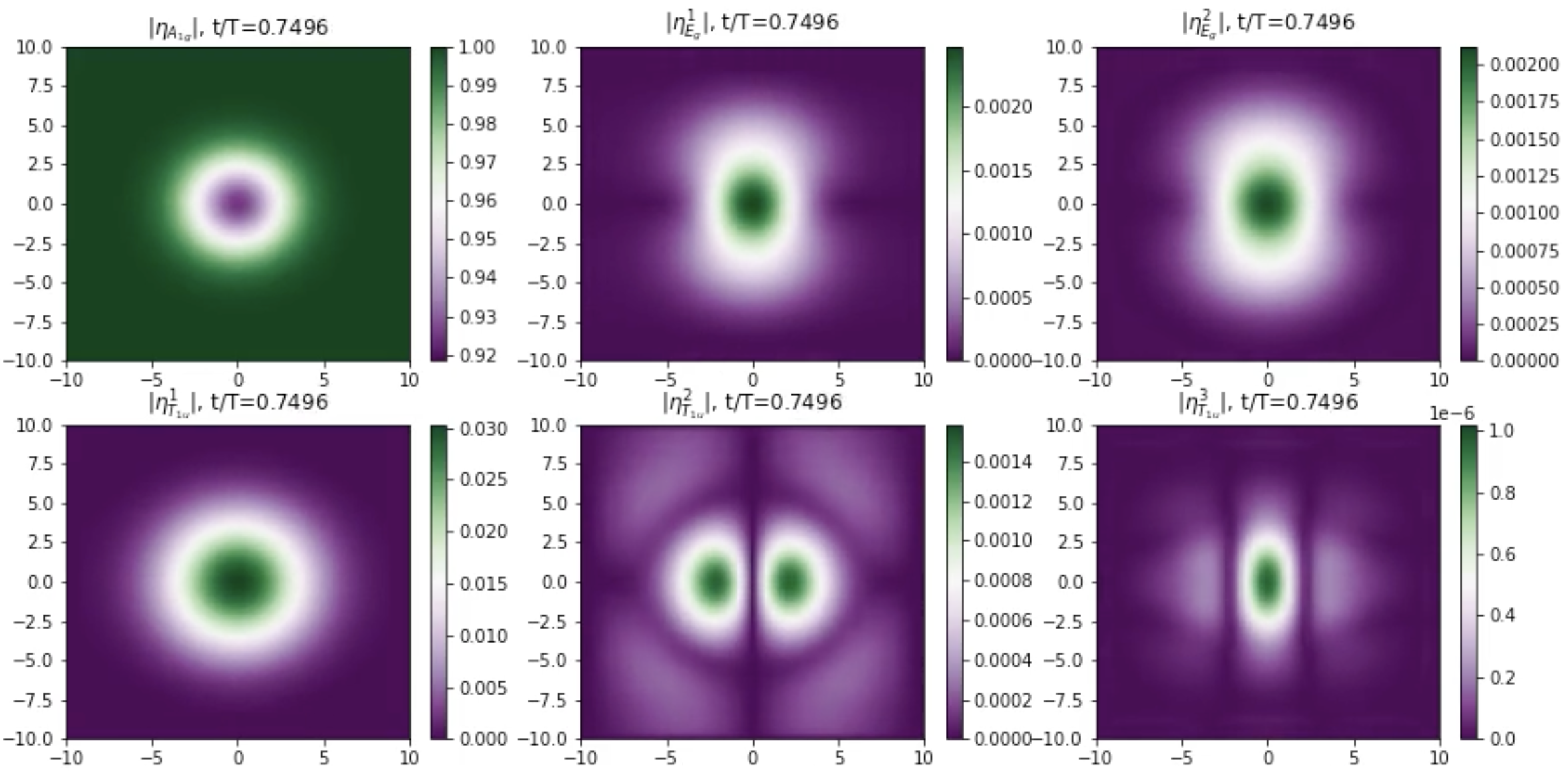}
    \caption{Density plots for the absolute values of $s$-wave $\eta_{A_{1g}}$, $d$-wave $\eta_{Eg}^{1,2}$, and $p$-wave $\eta_{T_{1u}}^{1,2,3}$ OPs under incident linearly polarized microwave beam at approximately quarter of the microwave cycle, $t/T\approx 0.75$, where $T$ is the period of microwave. Coordinates on the thin film are given in units of $\xi$.} 
\label{fig: Plots of OPs vs t linear3}
\end{figure*}

\begin{figure*}[!htbp]
    \includegraphics[width=\linewidth]{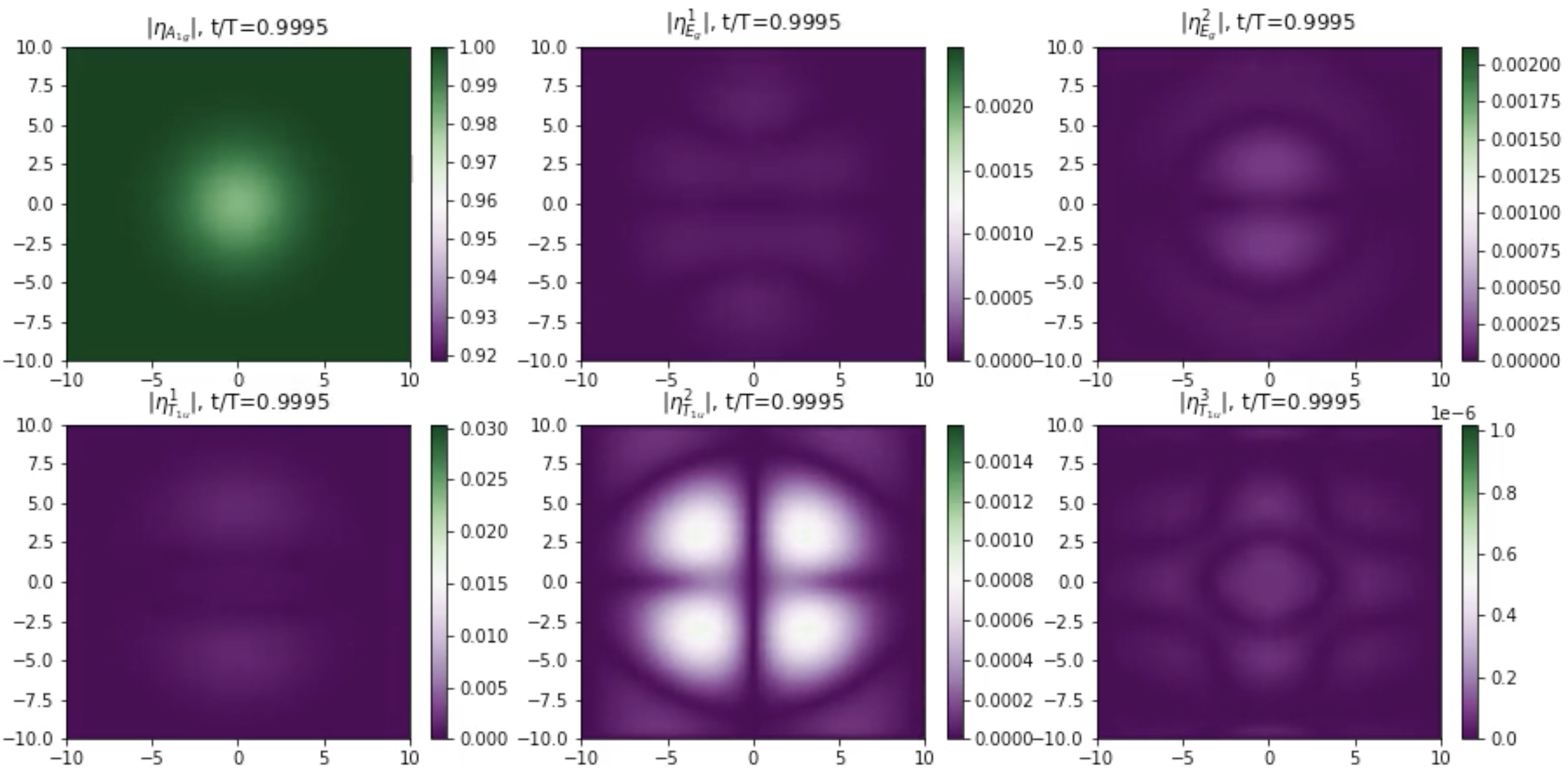}
    \caption{Density plots for the absolute values of $s$-wave $\eta_{A_{1g}}$, $d$-wave $\eta_{Eg}^{1,2}$, and $p$-wave $\eta_{T_{1u}}^{1,2,3}$ OPs under incident linearly polarized microwave beam at approximately quarter of the microwave cycle, $t/T\approx 1$, where $T$ is the period of microwave. Coordinates on the thin film are given in units of $\xi$.} 
\label{fig: Plots of OPs vs t linear4}
\end{figure*}

\begin{figure*}[!htbp]
    \includegraphics[width=\linewidth]{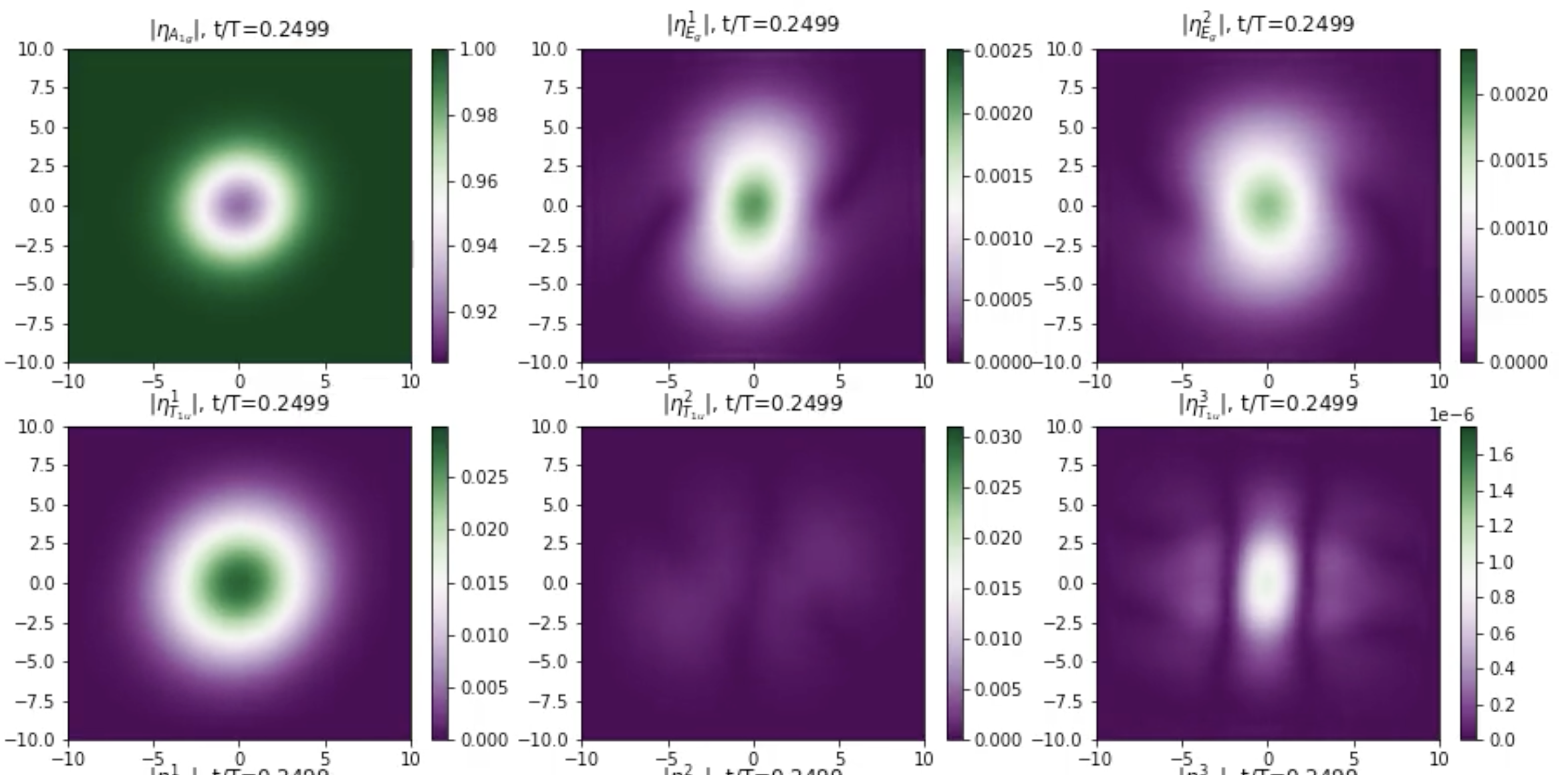}
    \caption{Density plots for the absolute values of $s$-wave $\eta_{A_{1g}}$, $d$-wave $\eta_{Eg}^{1,2}$, and $p$-wave $\eta_{T_{1u}}^{1,2,3}$ OPs under incident circularly polarized microwave beam at approximately quarter of the microwave cycle, $t/T\approx 0.25$, where $T$ is the period of microwave. Coordinates on the thin film are given in units of $\xi$.} 
\label{fig: Plots of OPs vs t circular1}
\end{figure*}

\begin{figure*}[!htbp]
    \includegraphics[width=\linewidth]{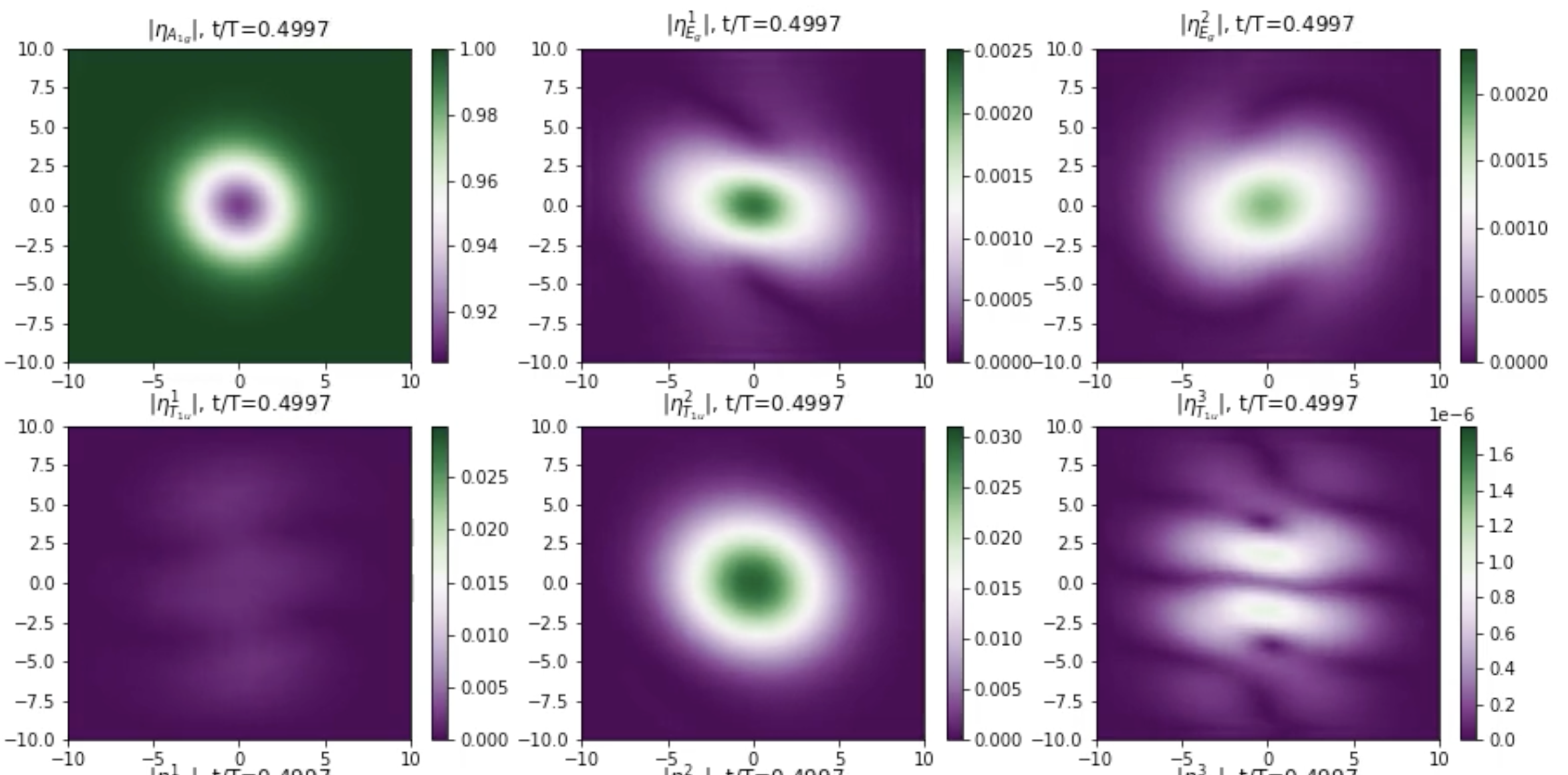}
    \caption{Density plots for the absolute values of $s$-wave $\eta_{A_{1g}}$, $d$-wave $\eta_{Eg}^{1,2}$, and $p$-wave $\eta_{T_{1u}}^{1,2,3}$ OPs under incident circularly polarized microwave beam at approximately half of the microwave cycle, $t/T\approx 0.5$, where $T$ is the period of microwave. Coordinates on the thin film are given in units of $\xi$.} 
\label{fig: Plots of OPs vs t circular2}
\end{figure*}

\begin{figure*}[!htbp]
    \includegraphics[width=\linewidth]{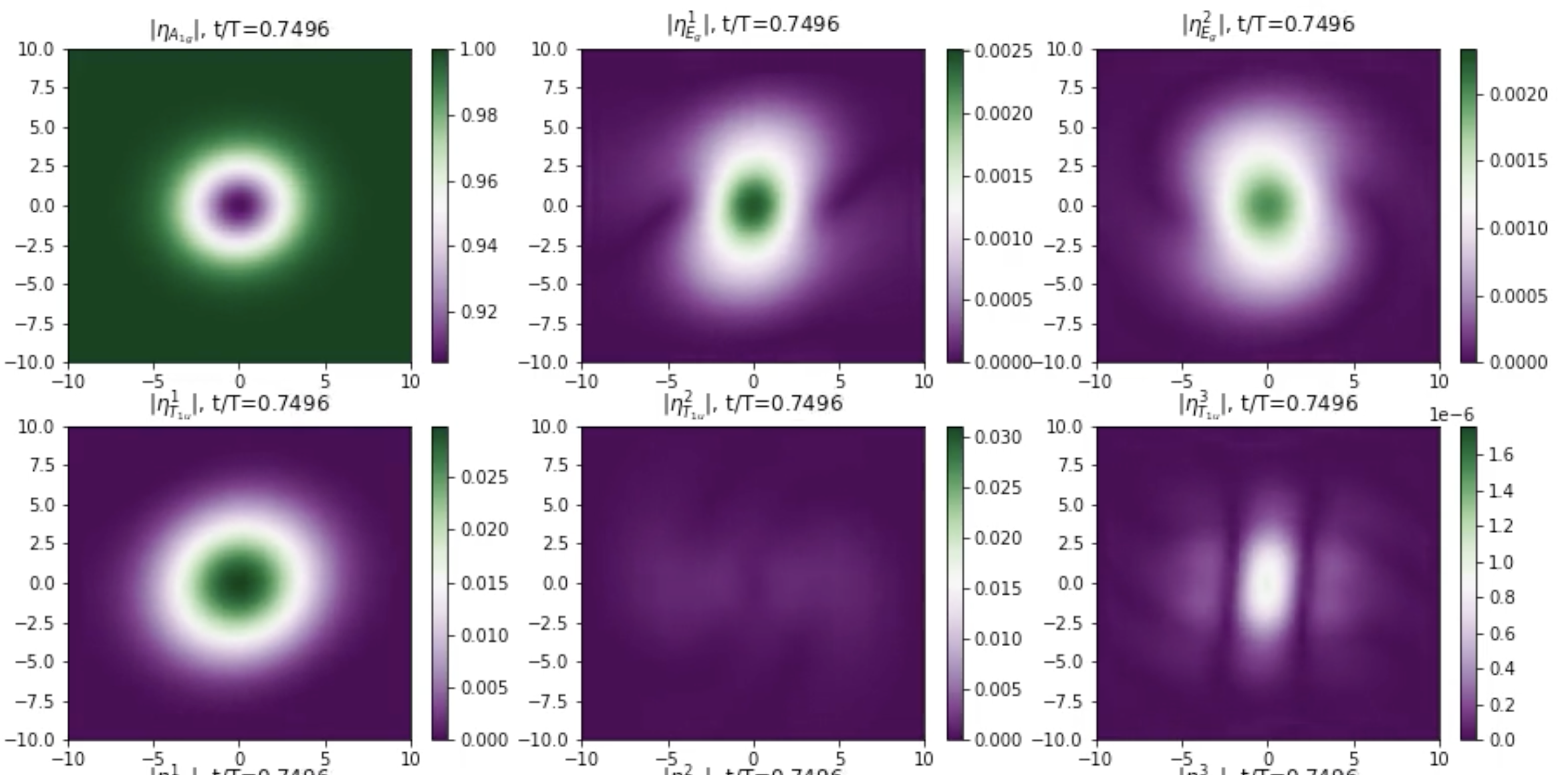}
    \caption{Density plots for the absolute values of $s$-wave $\eta_{A_{1g}}$, $d$-wave $\eta_{Eg}^{1,2}$, and $p$-wave $\eta_{T_{1u}}^{1,2,3}$ OPs under incident circularly polarized microwave beam at approximately quarter of the microwave cycle, $t/T\approx 0.75$, where $T$ is the period of microwave. Coordinates on the thin film are given in units of $\xi$.} 
\label{fig: Plots of OPs vs t circular3}
\end{figure*}

\begin{figure*}[!htbp]
    \includegraphics[width=\linewidth]{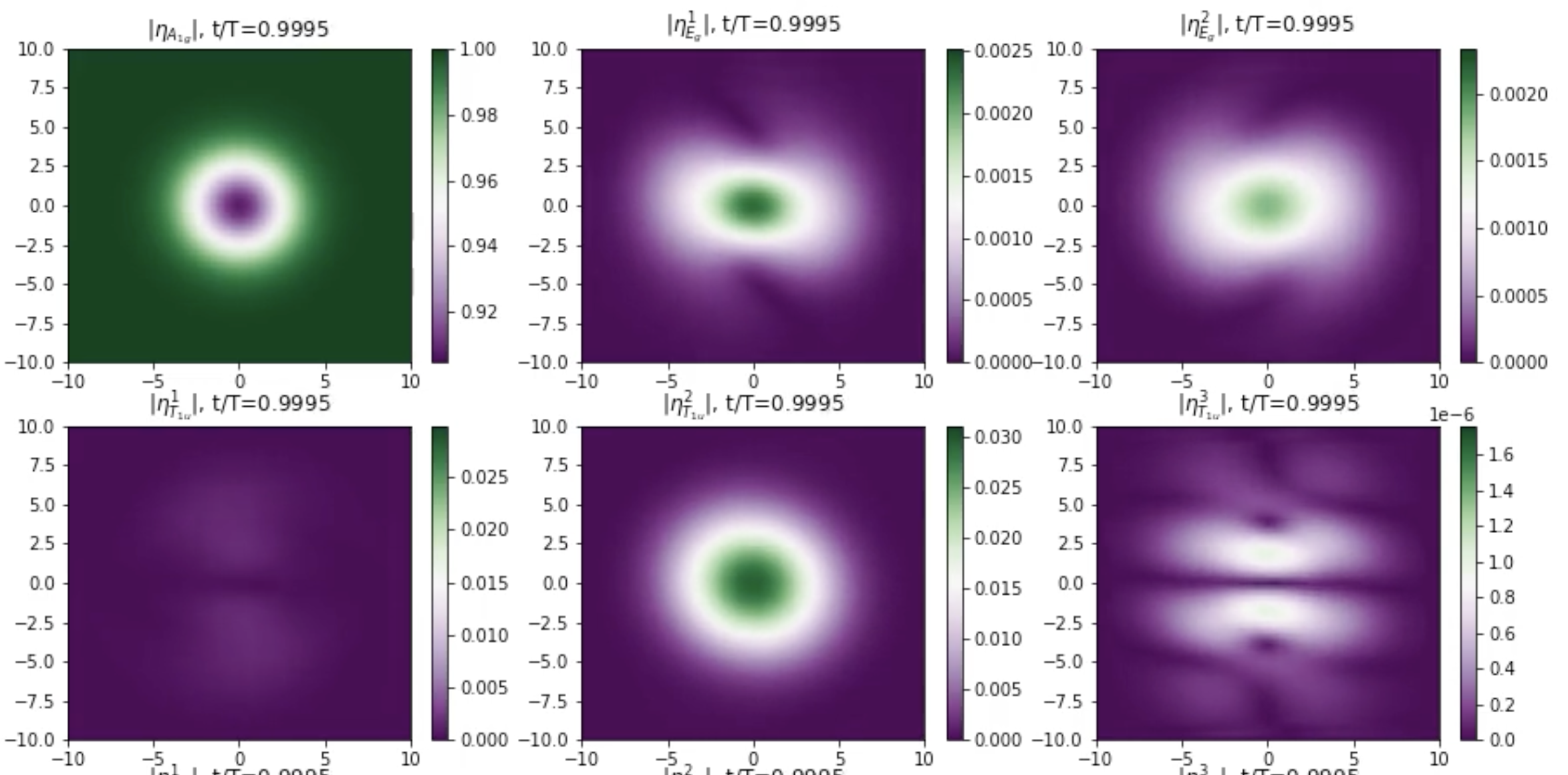}
    \caption{Density plots for the absolute values of $s$-wave $\eta_{A_{1g}}$, $d$-wave $\eta_{Eg}^{1,2}$, and $p$-wave $\eta_{T_{1u}}^{1,2,3}$ OPs under incident circularly polarized microwave beam at approximately quarter of the microwave cycle, $t/T\approx 1$, where $T$ is the period of microwave. Coordinates on the thin film are given in units of $\xi$.} 
\label{fig: Plots of OPs vs t circular4}
\end{figure*}

In this Appendix, we provide density plots for the absolute values of the $s$-wave $\eta_{A_{1g}}$, $d$-wave $\eta_{Eg}^{1,2}$, and $p$-wave $\eta_{T_{1u}}^{1,2,3}$ OPs under incident linearly and circularly polarized microwave beams at different times in~\cref{fig: Plots of OPs vs t linear1,fig: Plots of OPs vs t linear2,fig: Plots of OPs vs t linear3,fig: Plots of OPs vs t linear4} and~\cref{fig: Plots of OPs vs t circular1,fig: Plots of OPs vs t circular2,fig: Plots of OPs vs t circular3,fig: Plots of OPs vs t circular4}, respectively.

\bibliography{library}

\end{document}